\documentclass[a4paper,fleqn,usenatbib, useAMS]{mnras}
\usepackage{pdflscape}	% Landscape pages
\usepackage{txfonts}
\usepackage{natbib}
\usepackage{graphicx}
\usepackage{xspace}
\usepackage{longtable}
\usepackage{breqn}

\usepackage[T1]{fontenc}
\usepackage{ae,aecompl}
\setcitestyle{notesep={ }} %suppresses the second comma in \citep[][]{}
\newcommand{\ions}[2]{#1\textsc{#2}}

\title[Host galaxy environments of SLSNe-I]{Spatially resolved analysis of Superluminous Supernovae PTF~11hrq and PTF~12dam host galaxies}
\author[A. Cikota et al.]{Aleksandar Cikota$^{1}$\thanks{E-mail: acikota@eso.org},
Annalisa De Cia$^{1}$,
Steve Schulze$^{2}$,
Paul M. Vreeswijk$^{2}$, 
\newauthor Giorgos Leloudas$^{2,3}$,
Avishay Gal-Yam$^{2}$,
Daniel A. Perley$^{3}$,
Stefan Cikota$^{4,5}$,
\newauthor Sam Kim$^6$,
Ferdinando Patat$^{1}$,
Ragnhild Lunnan$^7$,
Robert Quimby$^{8,9}$,
Ofer Yaron$^{2}$,
\newauthor Lin Yan$^7$,
Paolo A. Mazzali$^{10,11}$
\\
$^{1}$European Southern Observatory, Karl-Schwarzschild-Str. 2, 85748 Garching b. M\"{u}nchen, Germany \\
$^{2}$Department of Particle Physics and Astrophysics, Faculty of Physics, Weizmann Institute of Science, Rehovot 76100, Israel\\
$^{3}$Dark Cosmology Centre, Niels Bohr Institute, University of Copenhagen, Juliane Maries Vej 30, 2100 Copenhagen, Denmark \\
$^{4}$University of Zagreb, Faculty of Electrical Engineering and Computing, Department of Applied Physics, Unska 3, 10000 Zagreb, Croatia\\
$^{5}$Ru{\dj}er Bo\v{s}kovi\'{c} Institute, Bijeni\v{c}ka cesta 54, 10000 Zagreb, Croatia \\
$^6$Instituto de Astrof\'{i}sica, Facultad de F\'{i}sica, Pontificia Universidad Cat\'{o}lica de Chile, Vicu\~{n}a Mackenna 4860, 7820436 Macul, Santiago, Chile \\
$^7$Department of Astronomy, California Institute of Technology, MC 249-17, 1200 East California Blvd., Pasadena CA 91125, USA\\
$^8$Department of Astronomy, San Diego State University, San Diego, CA 92182, USA\\
$^9$Kavli IPMU (WPI), UTIAS, The University of Tokyo, Kashiwa, Chiba 277-8583, Japan\\
$^{10}$Astrophysics Research Institute, Liverpool John Moores University, IC2, Liverpool Science Park, 146 Brownlow Hill, Liverpool L3 5RF, UK\\
$^{11}$Max-Planck Institut f\"{u}r Astrophysik, Karl-Schwarzschild-Str. 1, 85748 Garching b. M\"{u}nchen, Germany
}
 
\date{Accepted XXX. Received YYY; in original form ZZZ}

\pubyear{2017}

\begin{document}
\label{firstpage}
\pagerange{\pageref{firstpage}--\pageref{lastpage}}
\maketitle

\begin{abstract}
Superluminous supernovae (SLSNe) are the most luminous supernovae in the universe. They are found in extreme star-forming galaxies and are probably connected with the death of massive stars. One hallmark of very massive progenitors would be a tendency to explode in very dense, UV-bright, and blue regions. In this paper we investigate the resolved host galaxy properties of two nearby hydrogen-poor SLSNe, PTF~11hrq and PTF~12dam. For both galaxies \textit{Hubble Space Telescope} multi-filter images were obtained. Additionally, we performe integral field spectroscopy of the host galaxy of PTF~11hrq using the Very Large Telescope Multi Unit Spectroscopic Explorer (VLT/MUSE), and investigate the line strength, metallicity and kinematics. 
Neither PTF~11hrq nor PTF~12dam occurred in the bluest part of their host galaxies, although both galaxies have overall blue UV-to-optical colors. The MUSE data reveal a bright starbursting region in the host of PTF~11hrq, although far from the SN location. The SN exploded close to a region with disturbed kinematics, bluer color, stronger [\ions{O}{III}], and lower metallicity. The host galaxy is likely interacting with a companion. PTF~12dam occurred in one of the brightest pixels, in a starbursting galaxy with a complex morphology and a tidal tail, where interaction is also very likely. We speculate that SLSN explosions may originate from stars generated during star-formation episodes triggered by interaction. High resolution imaging and integral field spectroscopy are fundamental for a better understanding of SLSNe explosion sites and how star formation varies across their host galaxies. 
\end{abstract}


%________________________________________________________________

\section{Introduction}
%% to avoid "! pdfTeX error (ext4): \pdfendlink ended up in different nesting level than \pd " unmark following: 
%\vspace{55 mm}
%%%
Superluminous supernovae (SLSNe) constitute a new class of SNe that are %at least 1.5 magnitudes more luminous than the most luminous 
more luminous than classical SNe \citep[$M\leq-21$ mag, see][for a review]{2012Sci...337..927G}. Some SLSNe show hydrogen in their spectra (SLSN-II) and are mostly powered by the interaction between the SN ejecta and the circumstellar medium \citep[CSM,][]{2014ApJ...788..154O, 2016arXiv160401226I}. On the other hand, H-poor SLSNe (SLSN-I) do not show H in their spectra \citep{2011Natur.474..487Q} and are less understood. They might contain hydrogen, but it would be almost completely ionized near maximum, and therefore most likely not be visible \citep{2016MNRAS.458.3455M}. \citet{2015ApJ...814..108Y} have observed late-time hydrogen emission in one H-poor SLSNe. One possible powering source could be the spin-down of a magnetar \citep{2010ApJ...717..245K,2013ApJ...770..128I,2013Natur.502..346N}. Another possibility is the interaction between the ejecta and an H-poor CSM \citep{2012ApJ...746..121C,2017ApJ...835...58V, 2016ApJ...829...17S}. A third scenario could be a pulsational pair-instability SN \citep{2007Natur.450..390W, 2016arXiv160808939W}, or pair-instability explosion of a very massive star \citep[with a core of $\geq50M_\odot$, e.g.][]{2009Natur.462..624G,2013MNRAS.428.3227D}. The latter is perhaps limited only to slowly-declining H-poor SLSNe, SLSNe-R, with a declining rate consistent with the radioactive decay of $^{56}$Ni and $^{56}$Co. 
Until recently, models predicted a long-rise in contradiction to observations \citep{2015MNRAS.452.1567C}, but \citet{2017MNRAS.464.2854K} showed that they can also be fast rising. The long rise times predicted by PISN models have been seen in at least one SLSN \citep{2016ApJ...831..144L}. However, spectroscopic models predict spectra that are different from observations \citep{2016MNRAS.455.3207J}. SLSNe-R are a potential sub-class of SLSNe-I. The existence and characterization of this class is discussed in \citet{2012Sci...337..927G} and De Cia et al. (in prep.).    

While the debate on the powering source of SLSN-I is ongoing, the general agreement is that these explosions are associated with the death of massive stars, given their total radiated energies \citep[$E>10^{51}$ ergs,][]{2012Sci...337..927G}. The study of their host galaxies has revealed that SLSN-I require extreme hosts, e.g. with low-metallicity ($Z\leq0.5Z_\odot$) and a high rate of star formation (SFR) for their low mass \citep{2011ApJ...727...15N,2014ApJ...787..138L, 2016ApJ...830...13P,2016arXiv160504925C, 2015MNRAS.451L..65T, 2016arXiv161205978S}, and strong emission lines \citep{2015MNRAS.449..917L}. However, \citet{2016ApJ...830...13P} argue that a high rate of star formation is not required, since there are many counterexamples of SLSNe with low SFRs, based on slit spectorscopy. High-resolution observations with \textit{Hubble Space Telescope (HST)} showed that the host galaxies of hydrogen-poor SLSNe are usually irregular dwarf galaxies with asymmetric morphology or multiple peaks \citep{2015ApJ...804...90L, 2016MNRAS.458...84A}.

Gamma-ray bursts (GRBs) are another energetic phenomenon likely associated with deaths of massive stars \citep{2003Natur.423..847H}. Their host galaxies show similarities to the hosts of SLSNe \citep{2014ApJ...787..138L}, although the latter are more extreme \citep{2014ApJ...797...24V, 2015MNRAS.449..917L, 2016ApJ...830...13P,2016arXiv161205978S}.
\citet{2006Natur.441..463F} and \citet{2016ApJ...817..144B} showed that GRBs are more concentrated in the brightest regions of their host galaxies. 
Similarly, if SLSNe are indeed associated with very massive stars, and their hosts are unusual for the young stellar content \citep{2015MNRAS.451L..65T,2015MNRAS.449..917L}, one natural expectation is that they may explode in blue, UV bright, regions of their host galaxies, which are the most recent regions of massive star formation. This may be particularly relevant for pair-instability driven SLSNe-R, which are believed to have the most massive progenitors among SLSNe. \citet{2015ApJ...804...90L} found that the H-poor SLSNe sites are correlated with UV light, with a light fraction distribution between GRBs and an uniform distribution (see their Figure 6). \\

In this paper we investigate the spatially resolved properties, in particular the color and light fraction, of two among the nearest hydrogen-poor SLSNe (SLSNe-I) host galaxies. We focus on the host galaxies of two SLSNe-R, for which we have obtained \textit{HST} multi-band imaging. Furthermore, we obtained integral field spectroscopy of one of the hosts with VLT/MUSE, and additionally investigate the line strength, metallicity and the kinematics at the SN location.\\

The paper is organized as following. In Sect. 2 and 3 we describe the targets, and observations and data reduction, respectively. In Sect. 4 the methods and data analysis are presented. We then discuss the results in Sect. 5, and present our conclusions in Sect. 6.

\section{The targets}
The SLSNe-R PTF~11hrq and PTF~12dam were discovered by the Palomar Transient Factory \citep[PTF,][]{2009PASP..121.1395L,2009PASP..121.1334R}. While \citet{2016ApJ...830...13P} and \citet{2015MNRAS.451L..65T} included some preliminary commentary on the morphology and properties of host galaxies of PTF~11hrq and PTF~12dam, we will provide a more detailed quantitative and spatially resolved analysis of these systems.

\subsection{PTF~12dam}

PTF~12dam is among the most nearby SLSNe-I known, and happened in SDSS J142446.21+461348.6 at a redshift of \textit{z} = 0.1073. The host galaxy of PTF~12dam has also been studied by \citet{2015MNRAS.451L..65T}, \citet{2015MNRAS.452.1567C}, and \citet{2016ApJ...830...13P}, who found a metallicity of $12+\log($O/H$)=8.0$ throughout the galaxy. This metallicity is low, but not the most extreme compared to other SLSNe host \citep[some H-poor SLSNe have $12+\log($O/H$)\lesssim 7.8$; see e.g.][]{2015MNRAS.449..917L}. They also observed an extended asymmetric halo or tidal tail. Furthermore, the host galaxy of PTF~12dam has one of the highest star-formation rates (SFRs) among SLSNe hosts. \citet{2015MNRAS.451L..65T} performed partially resolved long-slit spectroscopy probing the kiloparsec environment of the SN site and suggest that the progenitor of PTF~12dam is a $>$ 60 M$_\odot$ star, formed in a recent starburst.

\subsection{PTF~11hrq}

PTF~11hrq is the nearest among all known SLSNe-I, and is also presented in \citet{2016ApJ...830...13P}. It exploded in an uncatalogued galaxy at a redshift of \textit{z} = 0.057. The galaxy might be a disk galaxy, but does not show a clear structure, and is not particularly young or starbursting, as estimated from slit spectroscopy. \citet{2016ApJ...830...13P} determined a metallicity $12+\rm{log_{10}(O/H)} = 8.15 \pm 0.03$ from emission-line diagnostics, a dust extinction $A_V=0.29 \pm 0.07$ mag (which corresponds to $E(B-V) \sim 0.07$ mag) from the Balmer-decrement, and a dust-corrected star-formation rate of $0.196 \pm 0.04 M_{\odot}$ yr$^{−1}$ based on the H$\alpha$ emission. They measured a rest-frame equivalent width of [\ions{O}{III}]$\lambda5007\AA$ for this galaxy, $EW([\ions{O}{III}]) = 40 \pm 1 \AA$\footnote{The quoted error is the statistical uncertainty and is usually underestimated, because it does not include effects such as slit corrections or continuum absorption features.}.

\begin{figure*}
\includegraphics[trim=25mm 30mm 0mm 20mm, width=18cm, clip=true]{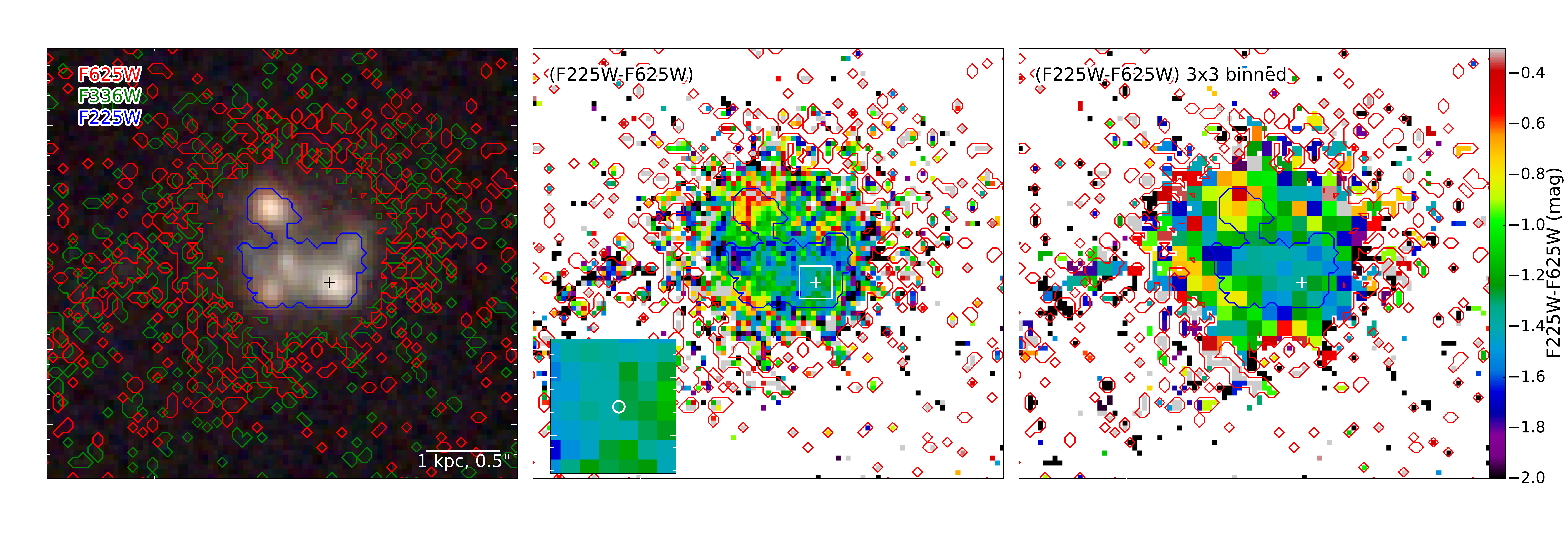} \\
\vspace{-2mm}
\caption{Color maps for the host galaxy of PTF~12dam. Left: A \textit{HST} composite color image. The contours denote the 3$\sigma$ flux level above noise for \textit{F225W} (blue), \textit{F336W} (green) and \textit{F625W} (red). Middle: \textit{F225W}-\textit{F625W} color map. The SN position is marked with the white cross. The region inside of the white box is zoomed-in in the subplot. The radius of the white circle in the inset plot corresponds to the position uncertainty of $\sim0.01$ arcsec. Right: color map of 3x3 binned data.}
\label{fig_ratio_12dam}
\end{figure*}

\begin{figure*}
\includegraphics[trim=25mm 30mm 0mm 20mm, width=18cm, clip=true]{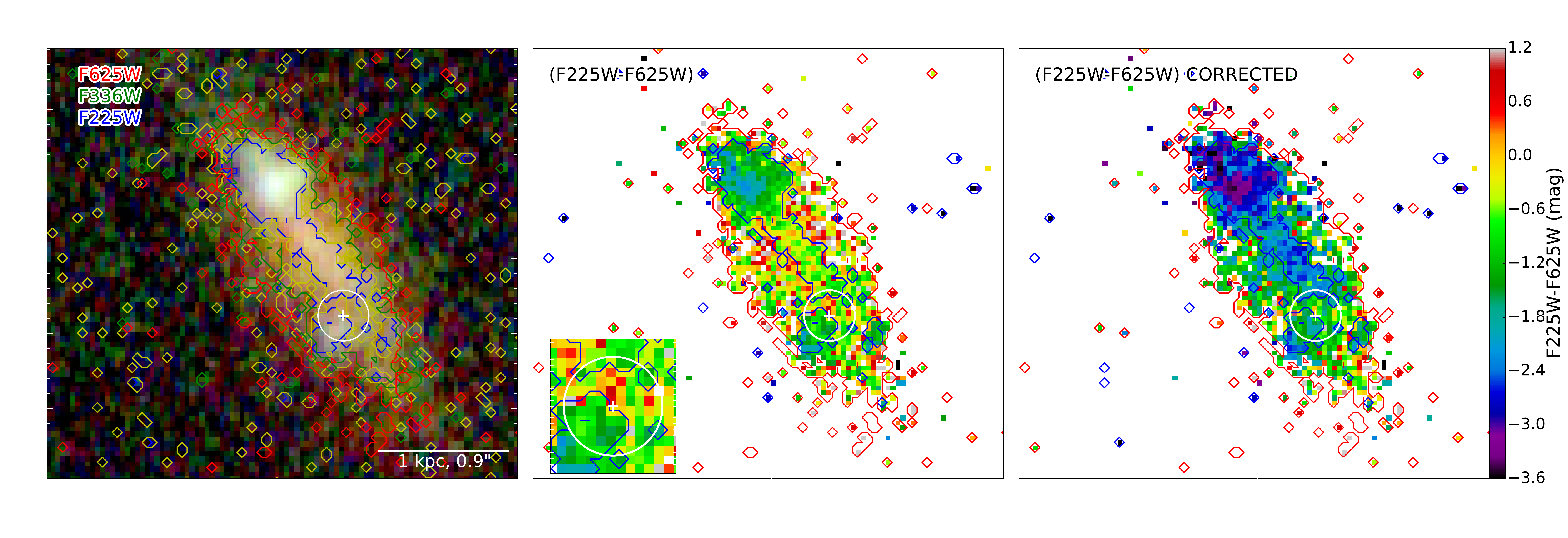} \\
\vspace{-2mm}
\caption{Color maps for the host galaxy of PTF~11hrq. Left: A \textit{HST} composite color image. The contours denote the 3$\sigma$ flux level above noise for \textit{F225W} (blue), \textit{F336W} (green) and \textit{F625W} (red). The yellow line denotes 2$\sigma$ flux level above noise for \textit{F225W}. Middle: \textit{F225W}-\textit{F625W} color map. The SN position is marked with the white cross. The radius of the white circle corresponds to the position uncertainty of $\sim0.17$ arcsec. Right: Color map corrected for dust extinction (see Section $\S$~\ref{extinciton}).}
\label{fig_ratio_11hrq}
\end{figure*}

\section{Observations and data reduction}
\label{sec:methods}
The host galaxies of PTF~12dam and PTF~11hrq were observed with \textit{HST} as part of the programme GO 13858 (PI: De Cia). They  were originally selected as host galaxies of SLSNe-R. The two galaxies were observed in Ultraviolet-Visible (UVIS) \textit{F625W}, \textit{F336W} and \textit{F225W} passband filters with the Wide Field Camera 3 (WFC3). A log of the observations is listed in Table~\ref{HSTlog}. 
Two more host galaxies of H-poor SLSNe (initially classified as SLSNe-R), PTF~13dcc \citep{2017ApJ...835...58V} and PTF~13ehe \citep{2015ApJ...814..108Y}, were observed as well as part of GO 13858, but only in one passband filter, \textit{F625W}, and at an early phase, while the SNe were still bright \citep{2013Natur.502..346N, 2017ApJ...835...58V}. Thus we do not investigate the host galaxies of PTF~13dcc and PTF~13ehe in this work.
The images were processed using standard astrodrizzle tools provided by the Space Telescope Science Institute (STScI), as described in more detail in $\S$3.2 of \citet{2016ApJ...830...13P}.

The effective wavelengths of the \textit{F225W}, \textit{F336W} and \textit{F625W} filters are 235.9 nm, 335.5 nm and 624.2 nm respectively, which correspond to restframe wavelengths of 213.0 nm, 303.0 nm and 563.7 nm for PTF~12dam at a redshift of \textit{z} = 0.1073; and 223.2 nm, 317.4 nm and 590.5 nm for PTF~11hrq (\textit{z}=0.057). \\

\begin{table}
\centering
\caption{Log of \textit{HST} observations}
%\vspace{-0.3cm}
\label{HSTlog}
\begin{tabular}{llcr}
\hline\hline
    Host galaxy &
    Observing date (UT) &
    Filter & 
    Exp. Time \\
    &&& (seconds) \\
\hline
  PTF~11hrq & 2014-11-14 05:04:49 & \textit{F625W} & 2$\times$100  \\
  PTF~11hrq & 2014-11-14 05:14:44 & \textit{F336W} & 2$\times$460 \\
  PTF~11hrq & 2014-11-14 06:39:48 & \textit{F225W} & 400+413 \\
  &&&\\
  PTF~12dam & 2014-10-16 17:59:45 & \textit{F625W} & 2$\times$100 \\
  PTF~12dam & 2014-10-16 18:09:40 & \textit{F336W} & 2$\times$492 \\
  PTF~12dam & 2014-10-16 18:42:09 & \textit{F225W} & 2$\times$433 \\
  \hline
\end{tabular}
\end{table}

Additionally, we observed the PTF~11hrq field with the Multi Unit Spectroscopic Explorer \citep[MUSE,][]{Bacon2010a}
integral field unit spectrograph mounted at UT4 of ESO's VLT. MUSE has a field-of-view of $60 \times 60$ arcsec, covered by 24 individual identical integral field units, each supplied with a slicer of 48 mirrors arranged in $4\times12$ stacks. Each spatial element (spaxel) has a size of $0.2 \times 0.2$ arcsec. This configuration gives a wide spectral range from 4750 to 9350 $\AA$ and a spectral resolution ranging from $R = 1800 - 3500$ across the wavelength domain. 

The observation of PTF~11hrq was obtained on October 1st, 2016 as a part of the programme 097.D-1054 (PI: Kim) and consisted of three exposures with an individual integration time of 1200 s. Each integration was dithered by 3 arcsec and rotated by 90$^\circ$. The full width at half maximum of the reconstructed $R$-band image (i.e. the seeing) is 0.71 arcsec.

We reduced the data in a standard fashion with the MUSE pipeline package version 1.6.2 that is a part of the ESO Recipe Execution Tool (ESOREX). We removed sky lines with the Zurich Atmosphere Purge (ZAP) software package \citep{Soto2016a} and matched the astrometry to our \textit{HST} images. 

\section{Methods and Data analysis}
\label{sec:analysis}

\subsection{Morphology of the host galaxies}
\label{morphology}

The \textit{HST} observations reveal a more complex morphology than what was accessible from ground-based imaging, for both host galaxies. In the case of PTF~12dam, the host galaxy has a complex morphology with five bright knots (Fig.~\ref{fig_ratio_12dam}). The extended asymmetric halo observed by \citet{2016ApJ...830...13P} and \citet{2015MNRAS.451L..65T} is partly visible in our images, as an extended structure in the west part of the galaxy. The host galaxy of PTF~11hrq has one bright knot in the northern part (Fig.~\ref{fig_ratio_11hrq}). 

\subsection{Supernovae position in their host galaxies}
\label{posdetermination}

\subsubsection{PTF~12dam position}

To locate the exact position of PTF~12dam in our \textit{HST} images, we use an \textit{HST} WFC3/UVIS image of the SN taken before maximum light, on May 26th, 2012 %at 01:52:58 UT 
(Proposal ID: 12524, PI: Quimby), through the \textit{F200LP} filter, and our \textit{F625W} image. Using SExtractor \citep{1996A&AS..117..393B} we extract sources on both images. The uncertainty on the barycenter position of the SN is less than a fraction of a pixel ($\sim$ 0.002 pixel, or $\sim$ 1$\times$10$^{-5}$ arcsec). We perform relative astrometry, using coordinate lists containing 6 common stars. We compute an image solution for our \textit{HST} data using the images.imcoords.ccmap routine from the PyRAF/STSDAS package.
The root-main-square (rms) of the image solution is 0.01 arcsecond for both right ascension and declination.
The final estimated position uncertainty of the SN (calculated by taking the square root of the sum of squared rms values) is $\sim$0.01 arcsec, and depends on the SN centroid uncertainty and the \textit{HST} image solution uncertainty. The SN is located between two pixels. Figure~\ref{fig_ratio_12dam} (left panel) shows a colour and countor-plot image of the host galaxy of PTF~12dam.

\subsubsection{PTF~11hrq position}

We determine the position of the SN in the \textit{HST} images relative to the Palomar Transient Factory (PTF) discovery image, observed on 2011-07-11. % at 11:27:16.511 UT.
Since there are no common stars in the \textit{HST} and PTF field, we use a VLT/FORS2 image observed on 2012-12-17 % at 01:17:38.401 UT 
(Prog ID: 090.D-0440(A), PI: Mazzali) as an intermediate image to transfer the world coordinate system (WCS) from the PTF to the \textit{HST} image. We localize the sources on all images using SExtractor. The uncertainty on the barycenter position of PTF~11hrq in the PTF field is $\sim$ 0.02 pixel, which corresponds to $\sim$ 0.02 arcsecond.
To confirm the position, we compared the PTF image from 2011-07-11 to PTF images from July 12, July 13, when the SN was at brightest. There is no discernible difference between the centroids in the images, aligned in WCS. The image from July 13 has a significantly larger PSF (seeing of 4.62 arcsec, compared to 3.46 arcsec on July 12 and 3.15 arcsec on July 11, as recorded in the fits header). 

We use 13 common stars in the PTF and FORS2 field, and compute an astrometric solution for the FORS2 field using the images.imcoords.ccmap routine from the PyRAF/STSDAS package. The rms of the fourth-order polynomial fits is 0.1 arcsecond in both axes, right ascension and declination.
There are only three common sources in the FORS2 and \textit{HST} fields, which is not sufficient to obtain an image solution, however, we calculated the rms deviation between the three common sources, and confirmed that the images are well aligned. The rms deviations are 0.12 and 0.13 arcseconds in right ascension and declination respectively.
The final estimated position uncertainty (calculated by taking the square root of the sum of squared rms values) of the SN is $\sim$ 0.17 arcsec, and consists of the alignment uncertainty between FORS2 and \textit{HST} image, the astrometry solution uncertainties of the FORS2 image, and the centroid's position uncertainty of the SN in the PTF image. Figure~\ref{fig_ratio_11hrq} (left panel) shows the color image and 3$\sigma$ contours of the host galaxy of PTF~11hrq.

To align the VLT/MUSE field to the HST astrometry, we use 11 common sources and compute an image solution using the images.imcoords.ccmap routine from the PyRAF/STSDAS package. The rms deviations were 0.18 and 0.16 arcsec in right ascension and declination respectively.
The final estimated position uncertainty of the SN in the MUSE field is $\sim 0.23$ arcsec in right ascension and 0.25 arcsec in declination.

\subsection{Color maps}
\label{ratiomaps}

We created $\textit{F225W}-\textit{F625W}$ and $\textit{F336W}-\textit{F625W}$ color maps to visualize the bluest regions in the galaxy, which are the regions with the most massive and hottest stars.

For resampling we used \textsc{SWarp} \citep{2002ASPC..281..228B}, which is a tool that uses the astrometric projection defined in the WCS to resample the images. In order to reduce the impact of noise and possible artifacts introduced during resampling, we also bin the images by $3 \times 3$ pixel, calculate the binned color maps and compare the results.
The color is calculated as $m_1-m_2=-2.5\log_{10}(f_1/f_2)$, where $f_1$ and $f_2$ are flux values in the corresponding filter passband filters.
The maps for PTF~12dam and PTF~11hrq are shown in Fig.~\ref{fig_ratio_12dam} and Fig.~\ref{fig_ratio_11hrq}, respectively.
Corrections for dust extinction are presented in Sect. \ref{extinciton}, and the results are discussed in Sect. \ref{discussion}.

\subsection{Light-distribution analysis}
\label{sec_lightdistribution}

Here we determine the fraction of light of the pixel in which the SN occurred, compared to the host galaxy light distribution. This technique was developed by \citet{2006Natur.441..463F} and allows us to quantify to which extent the SNe trace their host light distribution, regardless of the morphology. However, the light distribution depends on the filter passband and the definition of the galaxy boundaries, i.e. on the choice of which pixels are to be considered as part of the host galaxy.

To operate this choice in an objective way, we calculate the standard deviation of the sky background in two regions around the galaxies, and define a 3$\sigma$ threshold for each filter (see contours in Fig.~\ref{fig_ratio_12dam} and Fig.~\ref{fig_ratio_11hrq}). The result will strongly depend on the selected threshold. For instance, if the threshold is lower (e.g. 1$\sigma$), fainter pixels will be included, and the light fraction of the pixel where the SN occurred will be higher. 

\begin{figure}
\begin{center}
\includegraphics[trim=0mm 0mm 0mm 5mm, width=9cm, clip=true]{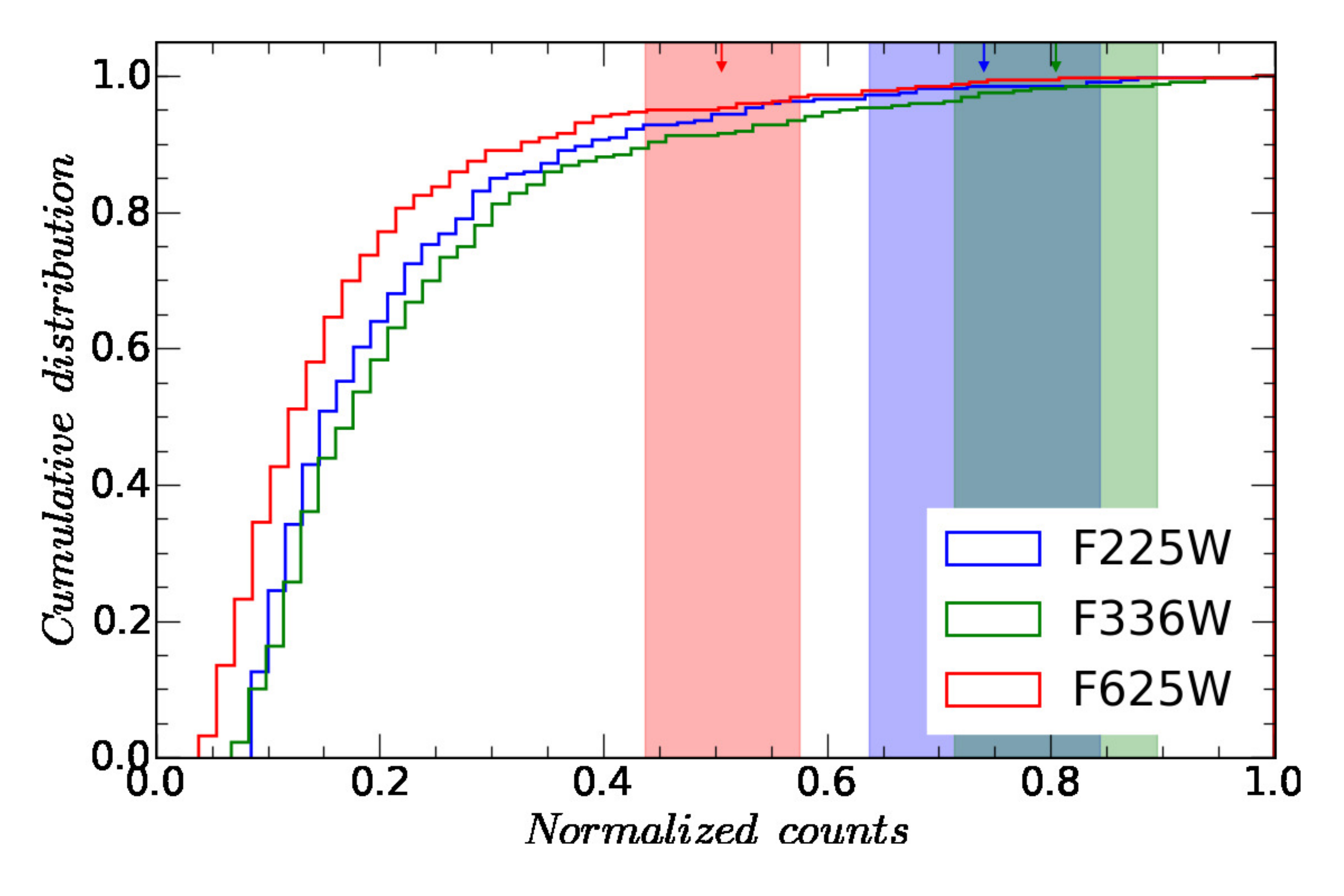}
\vspace{-6mm}
\caption{Cumulative histogram of light distribution in the host galaxy PTF~12dam in \textit{F625W} (red), \textit{F336W} (green) and \textit{F225W} (blue line). The arrows indicate the average count between the two pixels at the SN position, while the shaded areas are their standard deviations.}   %average between left and right filter
\label{fig_f_light_12dam}
\end{center}
\end{figure}

\begin{figure}
\begin{center}
\includegraphics[trim=0mm 0mm 0mm 5mm, width=9cm, clip=true]{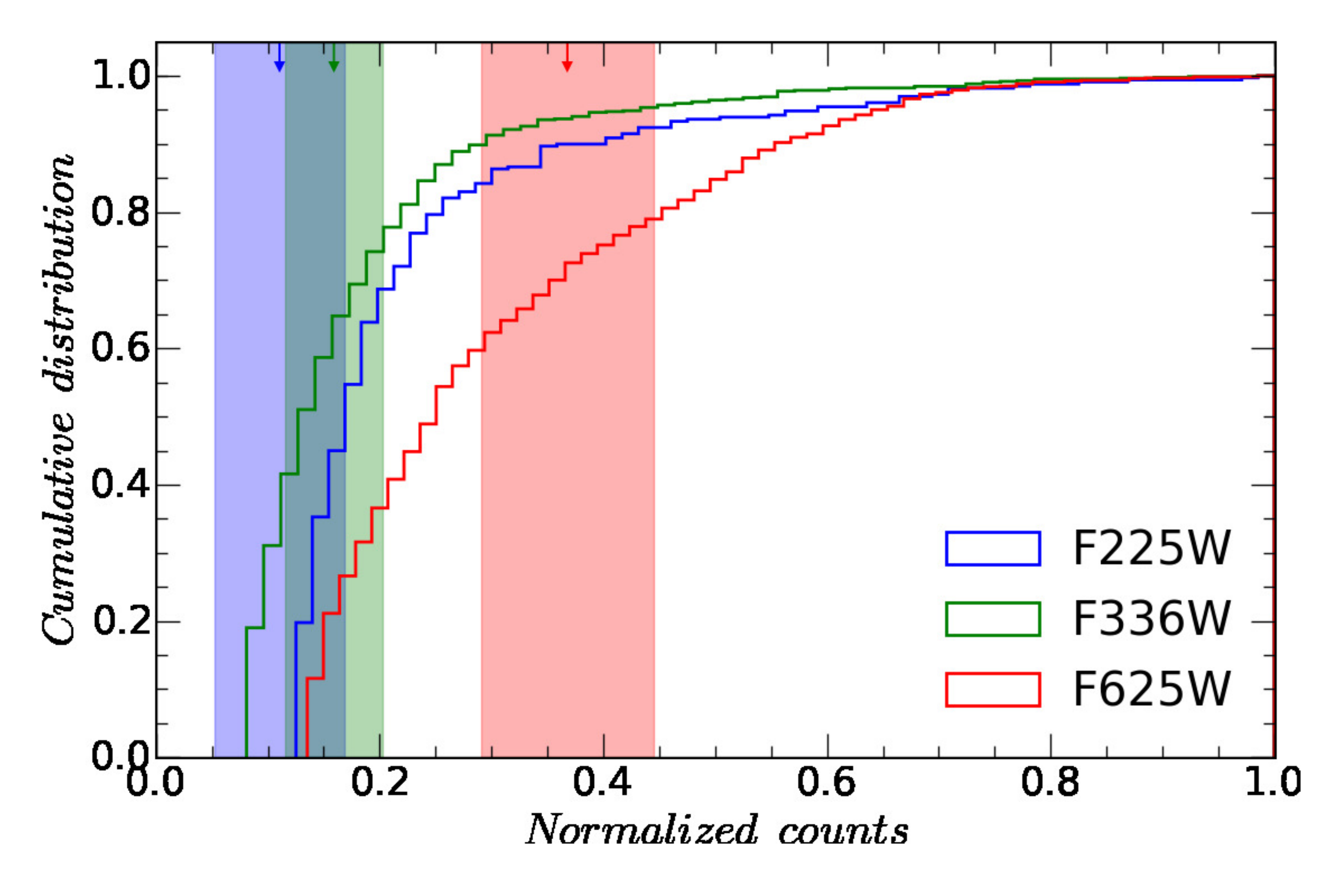}
\vspace{-6mm}
\caption{Cumulative histogram of light distribution in the host galaxy PTF~11hrq in \textit{F625W} (red), \textit{F336W} (green) and \textit{F225W} (blue line). The arrows indicate the mean counts within the 1$\sigma$ position uncertainty of the SN, while the shaded areas are their standard deviations.} 
\label{fig_f_light_11hrq}
\end{center}
\end{figure}

We measure the counts at the position of the SN, and calculate the light fractions from the cumulative distribution function. The results, for different thresholds and filters, are given in Table~\ref{tab:fraction}, and the cumulative distributions for the hosts of PFT 12dam and PTF~11hrq are shown in Fig.~\ref{fig_f_light_12dam} and Fig.~\ref{fig_f_light_11hrq} respectively.

We also calculate the light fractions using the intersection of the \textit{F225W}, \textit{F336W} and \textit{F625W} 3$\sigma$ thresholds as an universal threshold for all filters, with the aim to compare the light distribution of the different colors within the same region of the host.  
However, in the case of the host of PTF~11hrq, this biases the results against the older stellar populations, because part of the \textit{F336W} and \textit{F625W} fluxes are lost in the regions outside the universal threshold, which is in case of PTF~12dam  equivalent to the 3$\sigma$ \textit{F625W} threshold (Table~\ref{tab:fraction}).

\begin{table*}
\centering
\caption{Light distribution analysis of host galaxies.}
\vspace{-0.3cm}
\label{tab:fraction}
\begin{tabular}{lllll}
\hline\hline
Galaxy & Filter &  Threshold & No. of pixel & Light fraction at SN position \\
\hline
PTF~12dam & \textit{F225W} & 3$\sigma$ \textit{F225W} & 319  & 0.984 \\
          & \textit{F336W} & 3$\sigma$ \textit{F336W} & 2093 & 0.998 \\
          & \textit{F625W} & 3$\sigma$ \textit{F625W} & 2180 & 0.993 \\
  &  &  &   &  \\
PTF~12dam & \textit{F225W} & 3$\sigma$ \textit{F225W}$\cap$\textit{F336W}$\cap$\textit{F625W} & 319  & 0.984 \\
          & \textit{F336W} & 3$\sigma$ \textit{F225W}$\cap$\textit{F336W}$\cap$\textit{F625W} & 319  & 0.981  \\
          & \textit{F625W} & 3$\sigma$ \textit{F225W}$\cap$\textit{F336W}$\cap$\textit{F625W} & 319  & 0.953  \\  
  &  &  &   &  \\
PTF~11hrq & \textit{F225W} & 3$\sigma$ \textit{F225W} & 329 & 0.0 \\
          & \textit{F336W} & 3$\sigma$ \textit{F336W} & 837 & 0.648 \\
          & \textit{F625W} & 3$\sigma$ \textit{F625W} & 1184 & 0.723 \\
  &  &  &   &  \\
PTF~11hrq & \textit{F225W} & 3$\sigma$ \textit{F225W}$\cap$\textit{F336W}$\cap$\textit{F625W} & 286 & 0.0 \\
          & \textit{F336W} & 3$\sigma$ \textit{F225W}$\cap$\textit{F336W}$\cap$\textit{F625W} & 286 & 0.220  \\
          & \textit{F625W} & 3$\sigma$ \textit{F225W}$\cap$\textit{F336W}$\cap$\textit{F625W} & 286 & 0.311  \\      
  \hline
\multicolumn{5}{l}{\textbf{Notes.} \textit{F225W}$\cap$\textit{F336W}$\cap$\textit{F625W} is the intersection between all filters.}  \\
\end{tabular}
\end{table*}

\subsection{Analysis of the VLT/MUSE data of the PTF~11hrq host galaxy}
\label{musesection}

\begin{figure*}
\begin{tabular}{l}
%%%\hspace{-1.5cm}
\includegraphics[trim=12mm 0mm 0mm 22mm, width=18cm, clip=true]{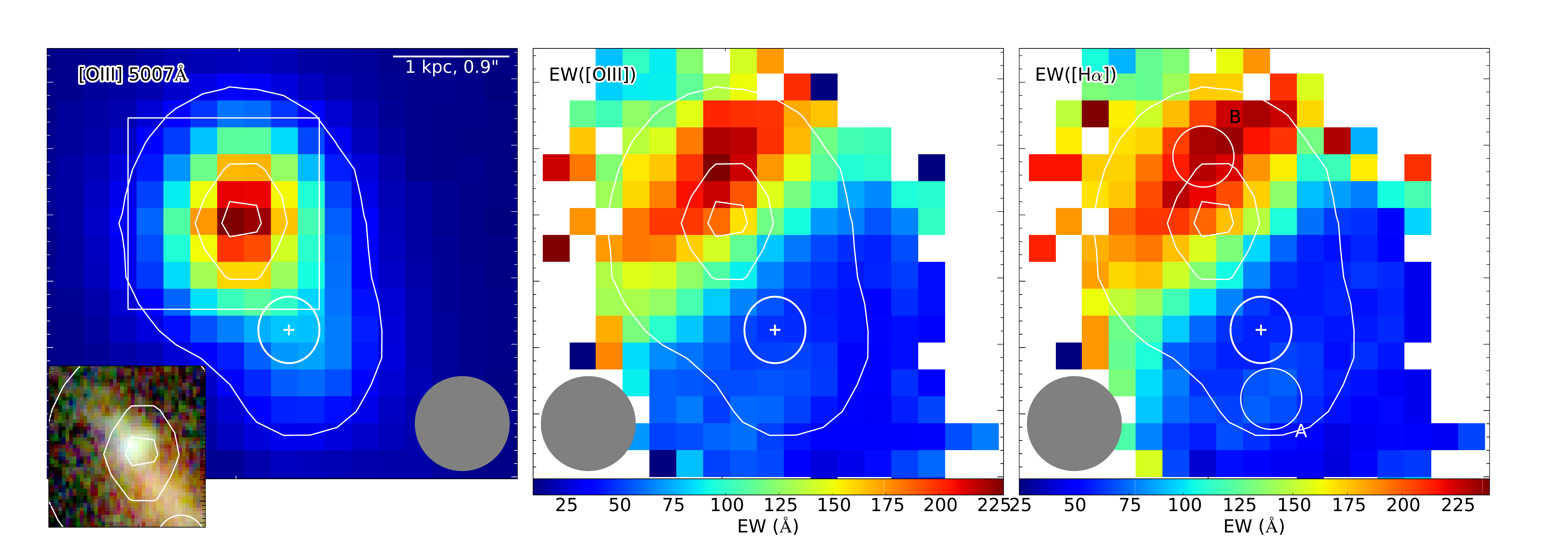} \\
%\hspace{-0.5cm}
%\vspace{-10mm}
\includegraphics[trim=12mm 0mm 0mm 22mm, width=18cm, clip=true]{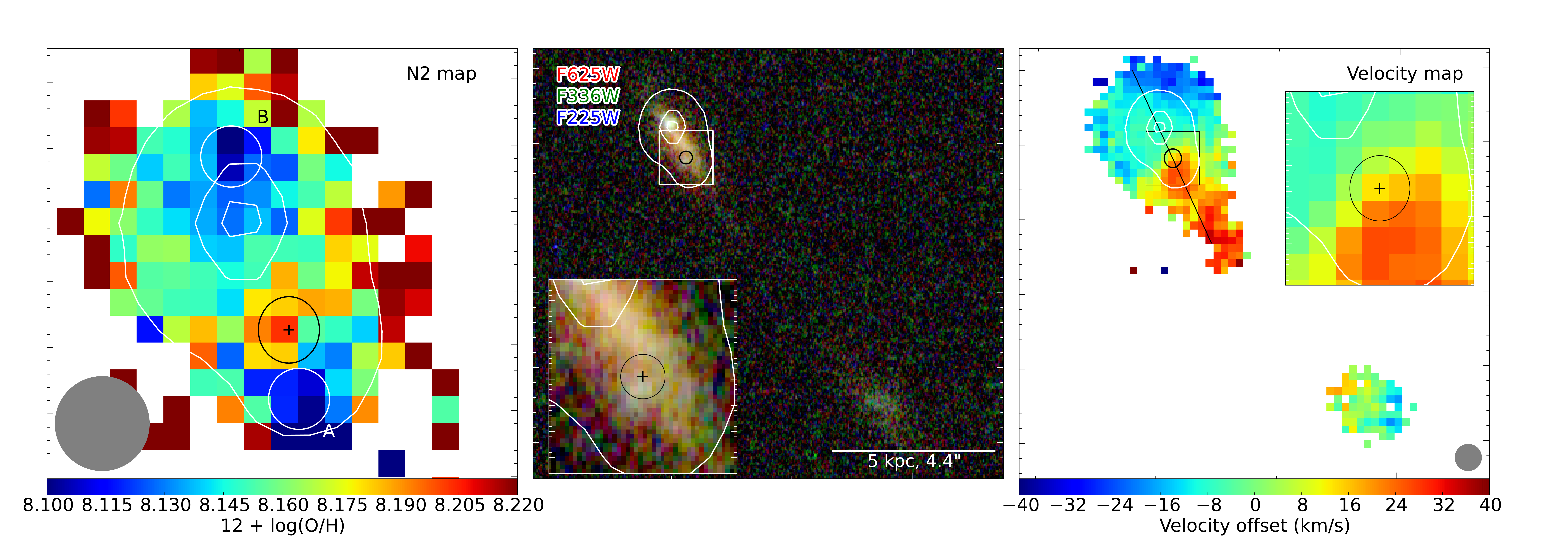} \llap{\makebox[63mm][l]{\raisebox{7mm} {\includegraphics[trim=10mm 5mm 5mm 5mm, width=3.1cm, clip=true]{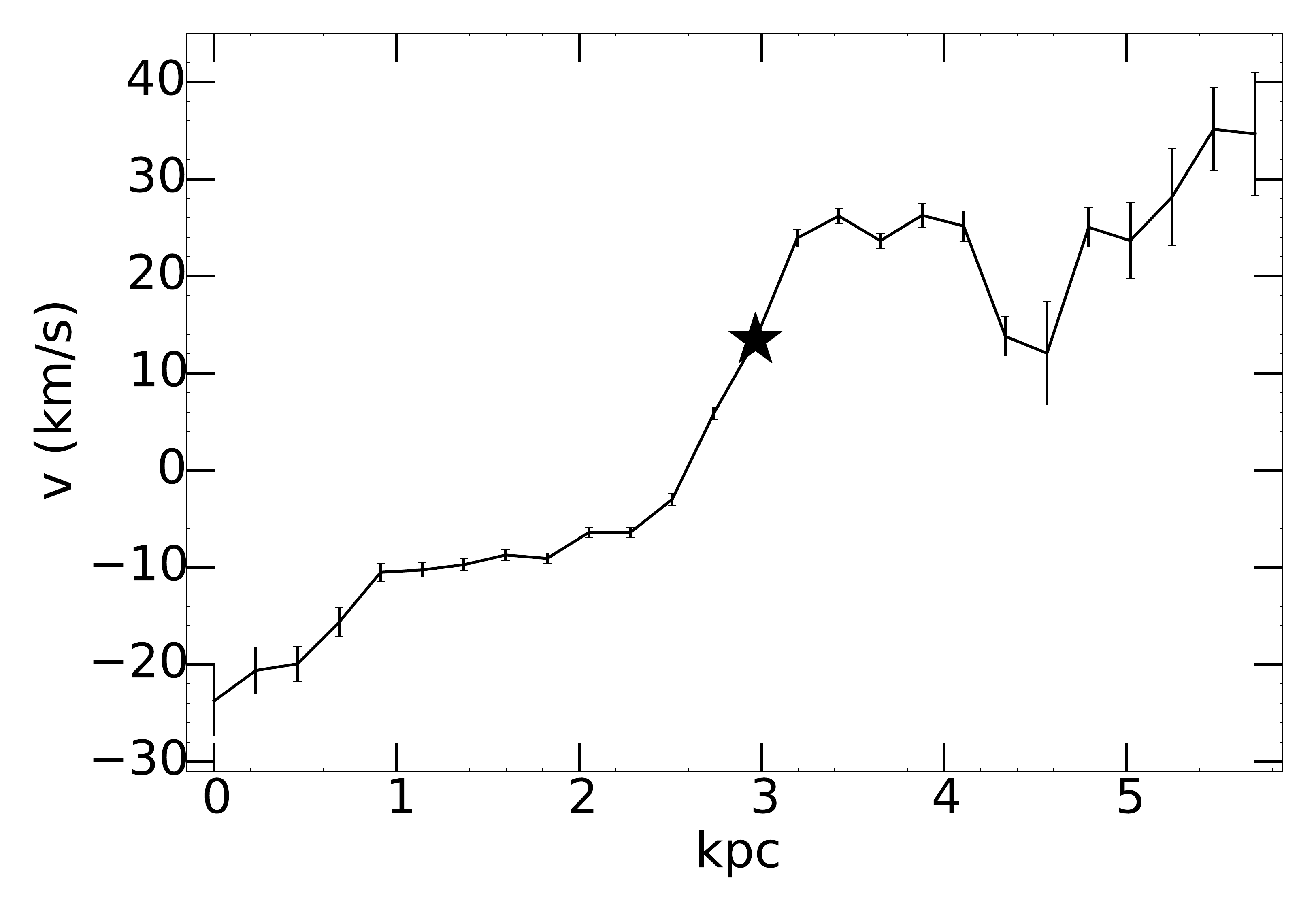}}}}\\ 
\end{tabular} 
\vspace{-2mm}
\caption{The host galaxy of PTF~11hrq. North is up, east is left. The SN location is marked with the '+' sign, and its position uncertainty is denoted with the circle for the \textit{HST} image and ellipse for MUSE. The seeing (0.71 arcsec) is marked with the gray circle in the bottom corner. Top left: [\ions{O}{III}]5007$\AA$ narrow band image. The [\ions{O}{III}] flux level contours are shown to make the comparison with other panels easier. The subframe is an \textit{HST} image with overplotted MUSE contours. Top middle: Equivalent width map of the [\ions{O}{III}]5007$\AA$ narrow emission line. Top right: equivalent width (EW) map of H$\alpha$. Bottom left: A metallicity map computed in the N2 scale using the \citet{2013A&A...559A.114M} calibration.  A and B are regions with lowest metallicities. Bottom middle: \textit{HST} image of the host galaxy. Bottom right: velocity map. The line indicates the position of the cut, shown in the inset frame. The star in the inset frame indicates the position of the SN.}
\label{fig_MUSE_11hrq}
\end{figure*}

\subsubsection{Metallicity map}
\label{sec_metallicity}

We compute the fluxes and equivalent widths (EW) of the H$\alpha$, H$\beta$, [\ions{O}{III}] and [\ions{N}{II}] emission lines. Then we compute metallicity maps from both the O3N2 and N2 index, using the \citet[][]{2004MNRAS.348L..59P} (hereafter PP04) and the updated \citet[][]{2013A&A...559A.114M} (hereafter M+13) calibrations, where O3N2 $\equiv\log\lbrace$([\ions{O}{III}]$\lambda$5007/H$\beta$)/([\ions{N}{II}]$\lambda$6583/H$\alpha$)$\rbrace^2$ \citep{1979A&A....78..200A} and N2 $\equiv\log\lbrace$[\ions{N}{II}]$\lambda$6583/H$\alpha\rbrace$ \citep{2002MNRAS.330...69D}. 

Despite the stellar absorption being very small and does not affect the metallicities significantly, in this paper we use only the N2 scale to estimate the absolute metallicity, because it depends on H$\alpha$, which is less affected by stellar absorption than H$\beta$. A re-reduced stellar absorption corrected metallicity map will be presented and further analyzed in a sample paper of SLSN host galaxies observed with MUSE (Schulze et al., in preparation). Nevertheless, both, the O3N2 and N2 scale can map the relative differences in metallicity throughout the host galaxy. 

The results and comparison between the PP04 and M+13 N2 indices are listed in Table~\ref{tab:metallicity}, and the metallicity map is shown in Fig.~\ref{fig_MUSE_11hrq}. The higher metallicity values at the edges of the galaxy have larger uncertainties, and should be considered with caution. The typical metallicity error is $\lesssim$ 0.15 dex (S/N of H$\alpha \gtrsim 5$), and metallicity values with an error > 0.24 dex (S/N $\lesssim 2$ in H$\alpha$) are ignored.

For comparison, also a background-subtracted [\ions{O}{III}] $\lambda$5007 $\AA$ line flux image is shown in Fig.~\ref{fig_MUSE_11hrq}. To build the [\ions{O}{III}] image, we defined a narrowband filter with a width of 16 $\AA$ and centred it at OIII, and to subtract the continuum, we applied a narrow-band filter with a width of 24 $\AA$ to the left and the right of the emission line and computed the mean value.

\begin{table*}
\centering
\caption{Local properties of the host galaxy of PTF~11hrq}
\vspace{-0.3cm}
\label{tab:metallicity}
\begin{tabular}{llllll}
\hline\hline
 Region & PP04 N2  & M+13 N2 & EW(H$\alpha$) [$\AA$]  & Stellar Age [Myr] & SFR [M$_{\odot}$ yr$^{-1}$]  \\
\hline
Galaxy$^{\mathrm{a}}$ & 8.17 $\pm$ 0.02 & 8.15 $\pm$ 0.03 & 130.9 $\pm$ 62.7 & 10.2$^{+1.3}_{-1.0}$ & 0.174 $\pm$ 0.009\\
SN position & 8.19 $\pm$ 0.01 &  8.18 $\pm$ 0.02 & 60.8  $\pm$ 2.0 & 11.6$^{+0.05}_{-0.00}$  & 0.007 $\pm$ 0.001\\
Region A & 8.13 $\pm$ 0.01 &  8.11 $\pm$ 0.01  & 70.7  $\pm$ 1.8 &  11.5$^{+0.02}_{-0.02}$ & 0.003 $\pm$ 0.001\\
Region B & 8.14 $\pm$ 0.01 &   8.12 $\pm$ 0.01 & 226.6  $\pm$ 7.2 &  8.9$^{+0.07}_{-0.1}$ & 0.016 $\pm$ 0.003\\
  \hline
\multicolumn{6}{l}{$^{\mathrm{a}}$ Mean within the outer contour in Fig.~\ref{fig_MUSE_11hrq}.} \\
\multicolumn{6}{l}{\textbf{Notes.} The stellar age errors are statistical errors derived from the EW(H$\alpha$) - age relation as shown in Fig.~\ref{fig:age}.}\\
\multicolumn{6}{l}{The SFR was converted from H$\alpha$ luminosity, as described in $\S$~\ref{appendix_SFR}.}\\
\end{tabular}
\end{table*}

\subsubsection{Stellar age}
\label{sec_stellar_age}

We estimated the age of the youngest ionizing stellar populations from the equivalent width of H$\alpha$, assuming a star formation law for an instantaneous burst, i.e. a single stellar population. \citet{1999ApJS..123....3L} provide equivalent width of H$\alpha$ as a function of stellar age in their Starburst99 model. The details of the calculation are presented in Appendix~\ref{appendix_sec_stellar_age}. 

The estimated stellar population ages in different regions of the host galaxy and at the SN position (see Fig.~\ref{fig_MUSE_11hrq}) are given in Table~\ref{tab:metallicity}.

\subsubsection{Velocity map}
\label{sec_velocity}

For each spatial pixel we fitted up to seven emission lines 
(H$\beta$ $\lambda$4862.68 $\AA$; [\ions{O}{III}] $\lambda$4960.295 $\AA$, $\lambda$5008.24 $\AA$; H$\alpha$ $\lambda$6564.61 $\AA$; [\ions{N}{II}] $\lambda$6585.27 $\AA$; [\ions{S}{II}] $\lambda$6718.29 $\AA$; [\ions{S}{II}] $\lambda$6732.67 $\AA$) using the MUSE Python Data Analysis Framework \citep[MPDAF,][]{2016ascl.soft11003B}, and created a redshift map by calculating the weighted average redshift of those lines. The redshifts with a weighted error larger than 3$\times 10^{-5}$ have been ignored. 

The average redshift of the galaxy is \textit{z} = 0.0569 $\pm$ 0.0001. The redshift map is shown in Fig.~\ref{fig_MUSE_11hrq}. South-west of the host galaxy, a fainter companion galaxy is visible at the same redshift, and at a projected distance of about 10 kpc. Furthermore, there is a third galaxy at a redshift of \textit{z}=0.0567, also in the south-west direction at a projected distance of 30 kpc, not visible in Fig.~\ref{fig_MUSE_11hrq}. This corresponds to a velocity difference of $\sim$45 km s$^{-1}$. Thus, these galaxies are likely part of the same group.

\subsubsection{Dust extinction correction}
\label{extinciton}

Normally one would expect young star forming regions to be characterized by blue colors. However, often they are the reddest parts of their galaxies, because of the large amounts of dust in which they are embedded. Although we do not expect much dust in low metallicity dwarf galaxies, dust can effectively attenuate UV radiation, and this could significantly affect the observed color. 

To correct the color map (Fig.~\ref{fig_ratio_11hrq}) for dust extinction, we derived the reddening from the Balmer decrement, and created a color excess map $E(\textit{F225W}-\textit{F625W})$ by adopting an extinction law, which we finally subtract from the observed color map. The typical value of $E(\textit{F225W}-\textit{F625W})_{\rm Host}$ is $\sim$ 0.8 $\pm$ 0.8 mag.
The calculations are discussed in more detail in Appendix~\ref{appendix_extinciton}.

\subsubsection{Star formation rate}
\label{appendix_SFR}

We convert the dust-corrected H$\alpha$ luminosities into star formation rates  via the relations presented in \citet{1998ARA&A..36..189K}, assuming a \citet{2003PASP..115..763C} initial mass function, i.e.:
\begin{equation}
{\rm SFR}_{{\rm H} \alpha } {\rm = 4.61 \times 10^{-42} L(H} \alpha {\rm) \hspace{10 pt} (erg\hspace{1 pt} s^{-1})}
\end{equation} 

\begin{figure}
\begin{center}
\includegraphics[trim=0mm 25mm 0mm 25mm, width=9cm, clip=true]{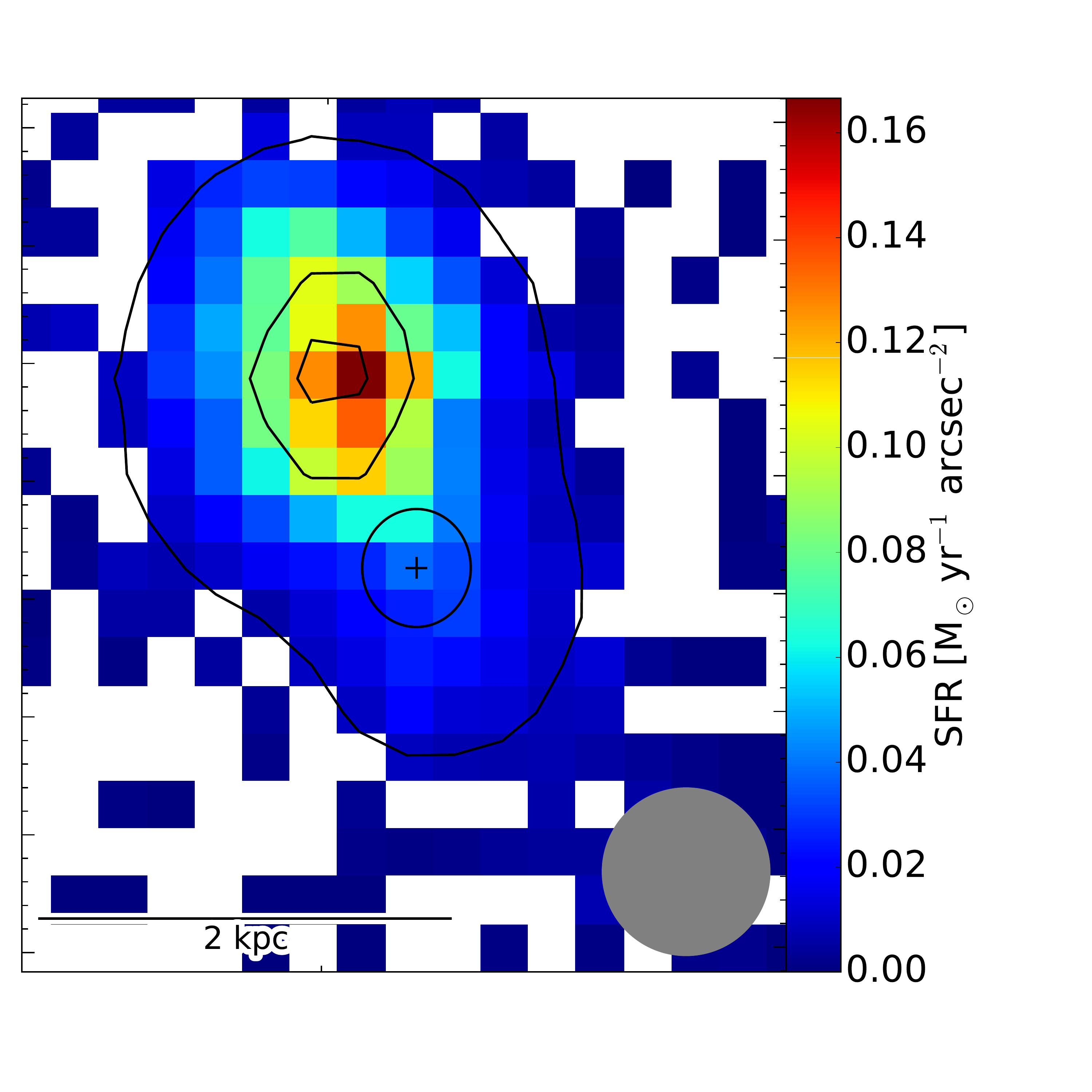}
\vspace{-4mm}
\caption{Star formation rate map for the host of PTF~11hrq. The contours are same as in Fig.~\ref{fig_MUSE_11hrq}. The black ellipse denotes the SN position uncertainty.}
\label{fig:sfr}
\end{center}
\end{figure}

The star formation rates (SFR) density map is shown in Fig.~\ref{fig:sfr}, and SFRs at different locations are listed in Table~\ref{tab:metallicity}. The sum of all values leads to a total SFR in of 0.174 $\pm$ 0.009 $M_{\odot}$ yr$^{-1}$, which is consistent within the errors with the SFR of $0.196 \pm 0.04 M_{\odot}$ yr$^{-1}$ determined by \citet{2016ApJ...830...13P}.

\section{Results and Discussion}
\label{discussion}

\begin{figure}
\begin{center}
\includegraphics[trim=0mm 0mm 0mm 0mm, width=9cm, clip=true]{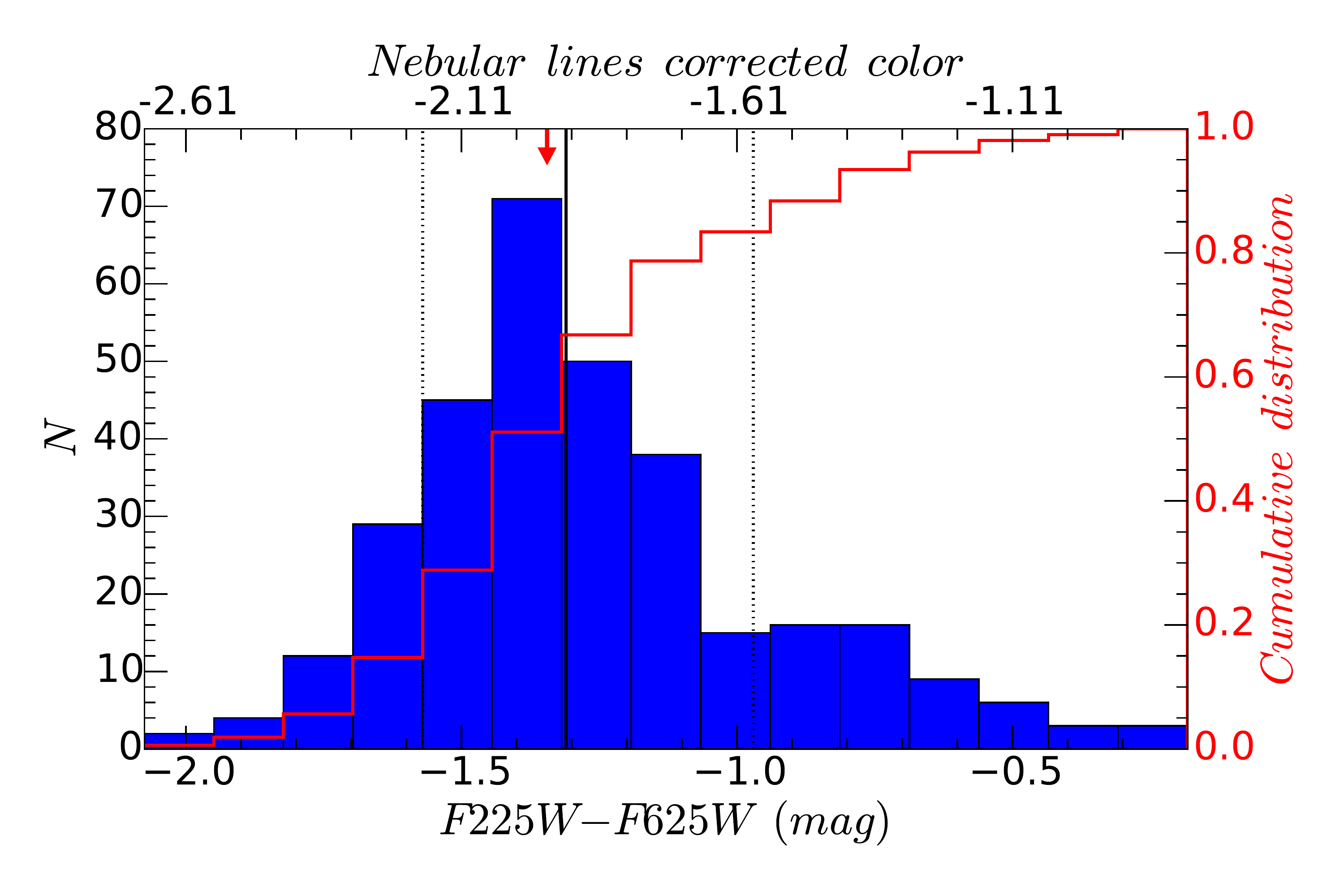}
\vspace{-6mm}
\caption{$\textit{F225W}-\textit{F625W}$ color distribution histogram of the host galaxy of PTF~12dam. Only values within the $3\sigma$ contours of the \textit{F225W} and \textit{F625W} image are taken into account. The solid black line is the mean of the distribution and the dotted lines shows its $1\sigma$ level deviations. The red arrow indicates the average color at the SN position. The color at the SN position is very close to the average value of the galaxy. The red line denotes the cumulative histogram. The upper x-axis shows the color after the nebular lines subtraction.} 
\label{fig_histo}
\end{center}
\end{figure}

\begin{figure*}
\begin{center}
\begin{tabular}{ll}
\includegraphics[trim=0mm 0mm 0mm 0mm, width=9cm, clip=true]{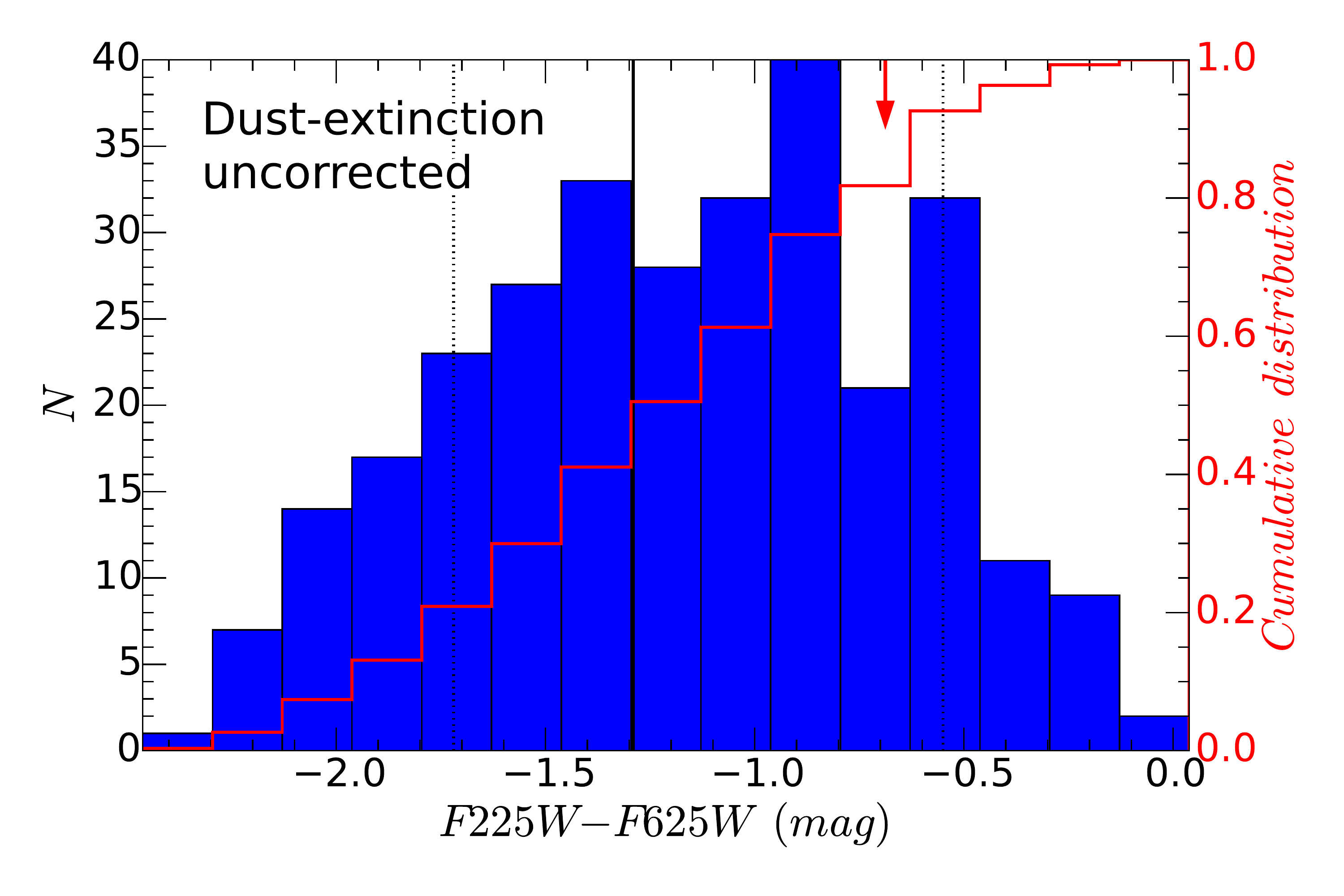}
&
\includegraphics[trim=0mm 0mm 0mm 0mm, width=9cm, clip=true]{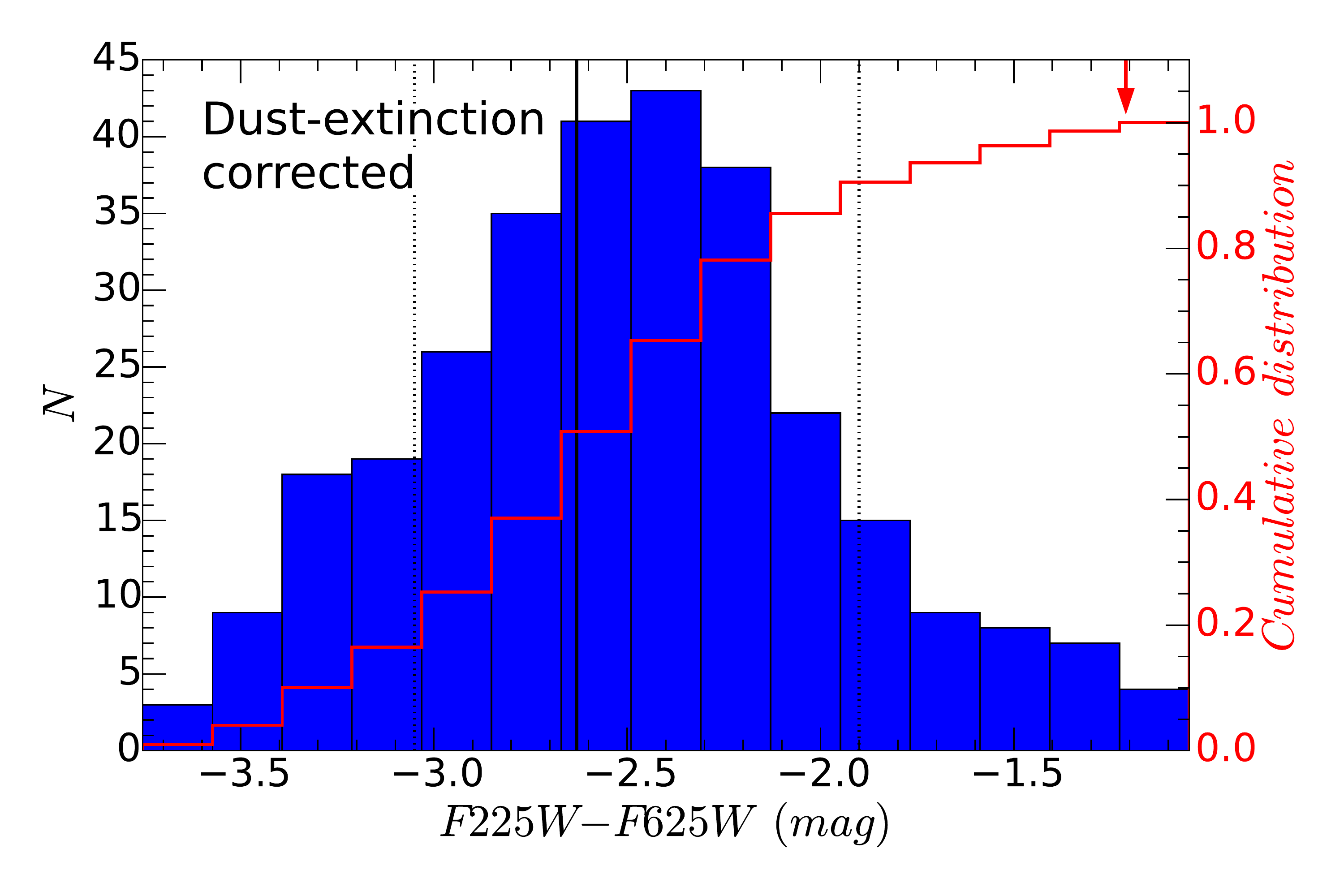}
\end{tabular}
\vspace{-4mm}
\caption{\textit{Left:} dust-extinction uncorrected $\textit{F225W}-\textit{F625W}$ color distribution histograms of the host galaxy of PTF~11hrq. Only values within the $3\sigma$ contours of the \textit{F225W} and \textit{F625W} image are taken into account. The solid black line is the mean of the distribution and the dotted lines are its $1\sigma$ levels. The red arrow marks the mean color within the uncertainty of the SN position. The red line denotes the cumulative histogram. 82$\%$ of pixels are bluer than the color at the SN position. \textit{Right:} dust-extinction corrected color distribution histogram. 93$\%$ of pixels are bluer than the color at the SN position.} 
\label{fig_histo_11hrq}
\end{center}
\end{figure*}

\begin{table*}
\centering
\caption{Colors of host galaxies environments}
%\vspace{-0.3cm}
\label{results}
\begin{tabular}{p{8cm}ll}
\hline\hline
           &   PTF~12dam &   PTF~11hrq \\
\hline
SN coordinates (J2000)\dotfill     & 14:24:46.207 +46:13:48.46$^{\mathrm{a}}$ & 00:51:47.251 -26:25:10.38$^{\mathrm{b}}$ \\
SN position uncertainty in \textit{HST} images \dotfill    & $\sim$ 0.01 arcsec &  $\sim$ 0.17 arcsec  \\
&&\\
$\textit{F225W}-\textit{F625W}$ color at SN position (mag)\dotfill & -1.40 (left pixel), -1.29 (right pixel) &  -0.69$^{\mathrm{e}}$ \\
$\textit{F225W}-\textit{F625W}$ color at SN position, 3x3 binned (mag)\dotfill &  -1.37 & -0.73$^{\mathrm{e}}$ \\
Mean $\textit{F225W}-\textit{F625W}$ color of galaxy (mag)$^{\mathrm{c}}$\dotfill &  -1.31 &   -1.29  \\
std  $\textit{F225W}-\textit{F625W}$ color of galaxy (mag)$^{\mathrm{c}}$\dotfill & $^{+0.34}_{-0.26}$ &  $^{+0.74}_{-0.43}$ \\
&&\\
$\textit{F336W}-\textit{F625W}$ color at SN position (mag)\dotfill & -0.69 (left pixel),  -0.57 (right pixel) &  -0.27$^{\mathrm{e}}$ \\
$\textit{F336W}-\textit{F625W}$ color at SN position, 3x3 binned (mag)\dotfill &  -0.67 &   -0.29$^{\mathrm{e}}$ \\
Mean $\textit{F336W}-\textit{F625W}$ color of galaxy (mag)$^{\mathrm{d}}$\dotfill &  -0.54  &  -0.42  \\
std $\textit{F336W}-\textit{F625W}$ color of galaxy (mag)$^{\mathrm{d}}$\dotfill  & $^{+0.38}_{-0.28}$ &  $^{+0.57}_{-0.37}$ \\
  \hline
  \multicolumn{3}{l}{ $^{\mathrm{a}}$ In the WCS of the \textit{HST} image taken on 2012-06-26 at 01:52:58.9 UT (Proposal ID: 12524).} \\        
  \multicolumn{3}{l}{ $^{\mathrm{b}}$ In the WCS of the PTF image taken on 2011-07-11 at 11:27:16.5 UT.} \\
  \multicolumn{3}{l}{ $^{\mathrm{c}}$ Mean color within the intersection of the 3$\sigma$ \textit{F225W} and \textit{F625W} contours.} \\
  \multicolumn{3}{l}{ $^{\mathrm{d}}$ Mean color within the intersection of the 3$\sigma$ \textit{F336W} and \textit{F625W} contours.} \\
  \multicolumn{3}{l}{ $^{\mathrm{e}}$ Mean within the uncertainty circle with an 0.17 arcsec radius.} \\
\end{tabular}
\end{table*}

We determined the positions of the SLSNe as described in $\S$~\ref{posdetermination}, calculated $\textit{F225W}-\textit{F625W}$ and $\textit{F336W}-\textit{F625W}$ color maps as described in $\S$~\ref{ratiomaps}, and performed a light distribution analysis ($\S$~\ref{sec_lightdistribution}). For the host galaxy of PTF~11hrq, which was also observed with VLT/MUSE, we investigated the metallicity ($\S$~\ref{posdetermination}), color excess $E$(H$\beta$-H$\alpha$) ($\S$~\ref{extinciton}), kinematics ($\S$~\ref{sec_velocity}) and determined the stellar age ($\S$~\ref{sec_stellar_age}).

PTF~12dam occurred in a dwarf galaxy which shows 5 bright knots. The SN position is in between two pixels in the \textit{HST} image (see Fig.~\ref{fig_ratio_12dam}), one pixel away ($\sim$ 68 pc)\footnote{For translating the angular distances into parsecs, we assume a flat universe with $H_0=67.8$ km s$^{-1}$ Mpc$^{-1}$ and $\Omega_M=0.308$ \citep{2016A&A...594A..13P}.} from the brightest pixel in the south-west knot of the galaxy. The light-distribution analysis in individual passband filters shows that the SN happened in one of the brightest pixels, at the 95th percentile (Table~ \ref{tab:fraction}, Fig.~\ref{fig_f_light_12dam}). The average $\textit{F225W}-\textit{F625W}$ color at the SN position is $-1.35$ mag, while the average color of the galaxy within the $3\sigma$ \textit{F225W} contours is $-1.31^{+0.34}_{-0.26}$ mag. This means that the color of the environment (i.e. the integrated color along the line of sight through the galaxy) at the SN position is average, deviating less than $0.35\sigma$ from the mean color value of the galaxy. We draw the same conclusions when binning the pixels by $3 \times 3$. The color at the SN position is at the 51st percentile, which means that 51$\%$ of pixels are bluer compared to the SN location (Fig.~\ref{fig_histo}). 

A caveat one should bear in mind when considering the color maps is that bright nebular lines can significantly contribute to the flux. The H$\alpha$ line indeed lies within the \textit{F625W} passband, and the host galaxy of PTF~12dam has strong H$\alpha$ emission. In Appendix ~\ref{appendix_emissionlinesimpact} we estimate the H$\alpha$ contribution and obtain that the UV-to-optical color is in fact 0.61 mag bluer, over the whole galaxy. The spatial distribution of the H$\alpha$, which could influence the relative colors inside the host, is however not known. Despite PTF~12dam exploded in an environment characterized by average colors of its host, this location is rather blue on an absolute scale ($\textit{F225W}-\textit{F625W} = -1.98$). The color distribution is shown in Fig.~\ref{fig_histo}, and the results are also summarized in Table~\ref{results}. 

For the host of PTF~12dam, the Galactic extinction is $E(B-V)_{\rm Gal} = 0.0107 \pm 0.0005$ mag \citep{2011ApJ...737..103S}. From spectral energy distribution modeling, \citet{2016arXiv161205978S} determined  a color excess of $E(B-V)_{\rm Host}\sim0.02$ mag, which is significantly lower than the reddening in the host of PTF~11hrq.

The host galaxy of PTF~11hrq has a bright peak visible in the \textit{F225W} image, in the northern part. Since the host galaxy is on average not particularly star-bursting \citep{2016ApJ...830...13P}, one might expect that the SN happened in the bluest and brightest star forming region, while it happened far away from this region, with a $5.5\sigma$ confidence. However, there is a second fainter blue knot just outside of the $\sim 1 \sigma$ uncertainty circle of the SN position (see Fig.~\ref{fig_ratio_11hrq}). The mean $\textit{F225W}-\textit{F625W}$ color within the SN position uncertainty circle is $-0.69$ mag, while the mean color values of the galaxy within the $3\sigma$ contours of \textit{F225W} and \textit{F625W} is $-1.29^{+0.74}_{-0.43}$ mag. For comparison, the $\textit{F225W}-\textit{F625W}$ color of the blue knot in the north is $-1.64^{+0.47}_{-0.33}$ mag. As shown in the color-distribution histogram (Fig.~\ref{fig_histo_11hrq}), the SN occurred in the red part of the galaxy. The results are summarized in Table~\ref{results}. However, the observed color is affected by dust reddening. The foreground Galactic reddening is $E(B-V)\sim0.01$ mag \citep{2011ApJ...737..103S}. We derived the host galaxy reddening from the VLT/MUSE data (see $\S$\ref{extinciton}), $E({\rm H}\beta-{\rm H} \alpha)_{Obs} = 0.13 \pm 0.12$ mag (Fig.~\ref{fig_Eba}), which corresponds to $E(B-V)\sim0.11$ mag, or $E(\textit{F225W}-\textit{F625W})_{Host} = 0.8 \pm 0.8$ mag. After correcting the color map for extinction, the color of the region where the SN occurred is $\textit{F225W}-\textit{F625W} \sim -1.2$ mag, at the $\sim$99th percentile (Fig.~\ref{fig_histo_11hrq}), i.e. among 1\% of the reddest pixels. 

The light-distribution analysis shows that PTF~11hrq did not happen in a bright pixel (Table~ \ref{tab:fraction}, Fig.~\ref{fig_f_light_11hrq}). In the F225W passband, the pixel counts at the position of the SN are below the $3\sigma$ threshold, and in F625W, depending on the definition of the threshold (see $\S$\ref{sec_lightdistribution}), $\sim72\%$ -- 31$\%$ of the pixels are brighter than the average value at the position of the SN. PTF~11hrq exploded far from the brighter and bluer region in the north part of its host galaxy, and therefore far from the region of likely stronger star formation. However, just outside of the uncertainty circle of the SN position, there is a second fainter blue region, particularly visible after the dust extinction correction, which might be evidence for a nearby star-forming region. 

The SLSN PTF~12dam occurred in one of the brightest \textit{F225W}, \textit{F336W} and \textit{F625W} pixels of an already very extreme galaxy. This is coherent with the findings of \citet{2015MNRAS.451L..65T}. They found that PTF~12dam occurred at a site of recent starburst with a very young stellar population ($\sim 3$ Myr), superimposed on an old stellar population. 

\citet{2015ApJ...804...90L} present a light fraction analysis of 16 hydrogen-poor SLSNe host galaxies in the rest-frame UV ($\sim 3000 \AA$) using \textit{HST}. PTF~12dam has a light fraction of 0.998 (in \textit{F336W}), which is at the $\sim$95th percentile compared to their sample \citep[see Fig. 6 in][]{2015ApJ...804...90L}, and PTF~11hrq has a light fraction of 0.65 (in \textit{F336W}), which is at the 56th percentile compared to their sample.\\

SLSNe may be associated with massive stars with zero-age main sequence masses of several hundred solar masses. Massive progenitors for SLSNe may be required to explain the large ejected masses derived from the light curves \citep[$M_{\rm ej}$ of 3--30 $M_\odot$,][]{2015MNRAS.452.3869N}. 
Therefore, it is expected that their location within the host galaxy is tightly correlated with the UV light. 
The evidence that both the SLSNe considered here did not explode in the bluest region of their host galaxies, where we would statistically expect them, is therefore surprising. In the case of recent starburst, there is an additional caveat. UV light probes star formation on a time scale of $\sim$ 100 Myr, while H$\alpha$ on time scales of $\sim$6 Myr \citep{2013seg..book..419C}. In case of a very young starburst, the region will still not be very UV bright. On the other hand, there is little evidence about the progenitor mass of SLSNe-I. Some models show that it is possible to get the spectral evolution for masses of less than 10 M$_{\odot}$ \citep{2016MNRAS.458.3455M}.

Given a sample of only two SLSNe, we cannot draw conclusions on the progenitors, based on the color of the environment. The progenitors of PTF~11hrq and PTF~12dam were not born in the most prominent blue regions of recent and massive star formation, but still, both galaxies are in general extreme. In particular, the host of PTF~11hrq has on average very young stellar population and low metallicity, with little variations between different regions. The host of PTF~12dam also has a young stellar population and a very high star-formation rate \citep[][]{2015MNRAS.451L..65T,2016ApJ...830...13P}. On an absolute scale, the two SN locations, and the two galaxies in general, are UV bright and blue. For instance, the $\textit{F225W}-\textit{F625W}$ color at the SN locations is -1.21 and $-1.98$ mag for PTF~11hrq and PTF~12dam, respectively, after the reddening and nebular-lines corrections.

Figure~\ref{fig_MUSE_11hrq} shows a comprehensive comparison between the [\ions{O}{III}] $\lambda$5007$\AA$ line flux image, [\ions{O}{III}] $\lambda$5007$\AA$ and H$\alpha$ equivalent width (EW) maps, N2 \citep{2013A&A...559A.114M} metallicity map, \textit{HST} image, and a velocity map. The peak of the [\ions{O}{III}] narrow band image coincides with the bright blue knot in the \textit{HST} image. %, which verifies the alignment between the HST and MUSE data. 
The EW([\ions{O}{III}]) and EW(H$\alpha$) maps show that the emission lines are stronger in the northern part of the galaxy, far from the SN location. The strongest lines, i.e. lines with largest equivalent widths, are 2--3 pixels north relative to the [\ions{O}{III}] intensity peak, with very high EW values up to $\sim 225 \AA$.

From EW(H$\alpha$), which comes from O stars exclusively, we estimate the stellar age in different regions of the host galaxy of PTF~11hrq (see $\S$~\ref{sec_stellar_age}). The average age of the galaxy is $\sim 10.2$ Myr, the youngest region is about 8.9 Myr old (Region B), while the SN occurred in a region of stellar age $\sim 11.6$ Myr (Fig.~\ref{fig:age}), which is not significantly different from the average.
%, and consistent with ages. 
Using the CMD 2.8 web interface\footnote{\url{http://stev.oapd.inaf.it/cgi-bin/cmd}} \citep{2012MNRAS.427..127B, 2014MNRAS.444.2525C, 2015MNRAS.452.1068C, 2014MNRAS.445.4287T}, from the stellar population age we estimate an upper limit on the stellar mass of $\sim 18 M_{\odot}$ at the SLSN location.

The N2 metallicity in the galaxy ranges from $12+\log({\rm O/H}) =\, \sim 8.05$--8.25, with an average (within the outer contour in Fig.~\ref{fig_MUSE_11hrq}) of 8.15 and a standard deviation of 0.06. Despite the metallicity range is rather narrow throughout the galaxy, there are two peaks of low metallicity. The minimum metallicity coincides with the region of the strongest emission lines (Region B), and south of the SN (Region A). The metallicity at the SN location is slightly, but clearly higher ($\sim 0.03$ dex) than the average (see Fig.~\ref{fig_MUSE_11hrq}). All metallicities are reported in Table~\ref{tab:metallicity}.

Remarkably, in Region A, south from the SN position uncertainty ellipse, just outside of the ellipse, there is a local increase of EW(H$\alpha$) and perhaps EW([\ions{O}{III}]), and a decrease in metallicity. 
Furthermore, the velocity map (bottom right panel of Fig.~\ref{fig_MUSE_11hrq}) reveals that there is an increase of velocity, with respect to the bulk of the galaxy, at this position. This suggests either locally disturbed kinematics in the galaxy, or a third small companion galaxy. The cut through the velocity map visualizes the galaxy rotation curve with the locally increased velocity near the position of the SN. The SN happened at the edge of the region with disturbed kinematics and is marked with a star in the velocity cut (Fig.~\ref{fig_MUSE_11hrq}). 
In summary, Region A, south of the SN uncertainty circle has a local peak in velocity, bluer $\textit{F225W}-\textit{F625W}$ color than average, stronger H$\alpha$ and [\ions{O}{III}] emission, and lower metallicity (Table~\ref{tab:metallicity}). These observations indicate likely past or ongoing interaction. 
This may suggest that local star formation has been triggered by interaction.
A more detailed study of the dynamics of the host galaxy is out of the scope of this paper, and will be presented in a paper about a sample of SLSN host galaxies observed with MUSE (Schulze et al., in preparation).
The host of PTF~12dam shows irregular morphology with multiple components in the \textit{HST} data, and a tidal tail \citep[see also][]{2015MNRAS.451L..65T,2016ApJ...830...13P}, perhaps hinting at past or ongoing interaction as well. 

Intriguingly, the MUSE cube of the PTF~11hrq host galaxy shows that it has two companions in the south-west, at projected distances of $\sim 10$ kpc and $\sim 30$ kpc. The nearby companion is at the same redshift as the host ($z = 0.0569$), with a velocity difference of $\sim$45 km s$^{-1}$. This suggests possible interaction. \citet{2003MNRAS.346.1189L} and \citet{2006MNRAS.367.1029S} found, based on analysis of galaxy pairs in the 2dF survey and SDSS, that galaxies with projected distances $r_p < 100 h^{-1}$ kpc, and relative velocities $\Delta V < 350$ km s$^{-1}$ have an enhanced star formation activity, induced by interaction. The companions around the host galaxy of PTF~11hrq fulfill those conditions.

Previous studies also suggest that interaction might play an important role in the formation of SLSN progenitors. For instance, \citet{2016arXiv161109910C} performed a detailed analysis of LSQ14mo, a type Ic SLSN likely powered by a magnetar spin-down, and its host galaxy system. They measured radial distances and velocities of three bright regions and propose that those are interacting components that have triggered star-formation. On the other hand, \citet{2016ApJ...830...13P} conclude that dynamical interaction or a specific SFR-dependent initial mass function is secondary, given that the majority of SLSNe-I in their sample occurred in host galaxies with SFRs typical of their low metallicity and stellar masses.

Interaction may trigger star formation in certain areas, such as the lower-density outskirts of galaxies, where other star-formation processes are less effective. Based on the velocity map of the host of PTF~11hrq, and the morphology of the host of PTF~12dam, we speculate that the supernovae originated from stars generated during recent star-formation episodes triggered by close interaction. Even though these areas may have overall less vigorous star formation (and appear redder) than other, perhaps denser, parts of these dwarf starbursting hosts, these areas may have specific conditions (for example, lower metallicity) that favor the formation of the massive progenitors of SLSNe. 

If SLSNe were associated with the most massive stars, we would statistically expect them to explode in the bluest and brightest starforming regions of their hosts. A larger statistical sample is needed to draw conclusions on the progenitors based on color. 

In this paper we studied the spatially resolved properties of two nearby SLSN host galaxies, using HST multi-band imaging and VLT/MUSE IFU spectroscopy. These reveal a complexity in morphology and spectral properties that was not accessible from ground-based imaging or slit spectroscopy. The complex morphology of PTF~12dam became evident only thanks to the HST resolution, which allowed a more detailed study of the star-formation regions in terms of color and light fraction. Furthermore, only with HST and MUSE we could appreciate the bright star forming region in the north part of the host of PTF~11hrq, and intensely emitting in H$\alpha$ and [\ions{O}{III}], as well as the presence of interacting companions. These features were missed, or smoothed away using classical slit spectroscopy. Further complexity in the morphology and spectral properties of more distant SLSN host galaxies in general can therefore be expected.

\section{Summary and conclusions}
\label{sec:summary}

We studied the host galaxies of two hydrogen-poor SLSN-I/R, PTF~12dam and PTF~11hrq, obtained with the \textit{Hubble Space Telescope} in \textit{F225W}, \textit{F336W} and \textit{F625W} passband filters, and compared the environment at the position of the SNe with the rest of the galaxy. Additionally, we obtained integral field spectroscopy with VLT/MUSE of the PTF~11hrq host galaxy. 

Our results can be summarized as follows:
\begin{enumerate}
\item HST observations of PTF~12dam reveal a complex and disturbed morphology with multiple star-forming knots. Combined with deep ground-based images, this complex represents the head of tadpole galaxy. The SN exploded in one of the five star forming knots. The explosion site is among the 5$\%$ of brightest pixels in the optical as well as the UV. While the UV-to-optical color appears rather average within the galaxy, it is blue ($-1.98$ mag, after correcting for nebular lines) and UV-bright on an absolute scale. 
\item The host galaxy of PTF~11hrq has a disk-like morphology and is characterized by a large population of evolved stars. HST and IFU observations reveal diverse unexpected features. The northern part of host contains a vigorously star-forming region with a [\ions{O}{III}] $\lambda$5007$\AA$ EW of >200 $\AA$. Intriguingly, the explosion site does not coincide with this region at $\gtrsim$5.5$\sigma$ confidence. 
The UV-to-optical color of the SLSN location is $-1.2$ mag, after correction for dust extinction, which is among 1\% of the reddest pixels, but nevertheless blue on an absolute scale and similar to PTF~12dam.  
Within the uncertainties, the SN location includes regions with higher metallicity and weaker [\ions{O}{III}] and H$\alpha$ emission lines relative to the bright part of the galaxy in the north. South of the SN position, at the edge of 1 sigma circle, there is a fainter blue knot which coincides with a region of locally lower metallicity, stronger H$\alpha$ emission and more disturbed kinematics.
\item The host galaxy of PTF 11hrq has a nearby ($\sim 10$ kpc) fainter companion at the same redshift, and possibly another one at a distance of about 30 kpc. The small angular separation and the small radial velocity difference of $\lesssim 40$ km s$^{-1}$ suggests that they could be part of a galaxy group. 
\item Both host galaxies show some evidence for interaction, and for PTF~11hrq this is possibly related to the SN explosion site. We speculate that the SLSN explosions may originate from stars generated in regions of recent or ongoing interaction.
\item The combination of high-resolution imaging and integral-field spectroscopy allowed for examining the conditions of the explosion sites and how star-formation varies across the host galaxies. Larger samples are needed to extract robust constraints on the progenitor population and how their galaxy environments affect the star formation process.
\end{enumerate}

\section*{Acknowledgements}
We are grateful to Wolfgang Kerzendorf (ESO Garching) for useful discussion and suggestions, in particular related to pixel alignment; and to Eliceth Rojas-Montes (Armagh Observatory) and Giacomo Beccari (ESO) for discussion related to massive stars, wind models and mass-loss rates.

STSDAS and PyRAF are products of the Space Telescope Science Institute, which is operated by AURA for NASA.

Based on observations made with ESO Telescopes at the Paranal Observatory under programme ID 090.D-0440(A) and 097.D-1054.

Based on observations made with the NASA/ESA Hubble Space Telescope, obtained at the Space Telescope Science Institute, which is operated by the Association of Universities for Research in Astronomy, Inc., under NASA contract NAS 5-26555. These observations are associated with program $\#$ 13858 and 12524.

This research made use of APLpy, an open-source plotting package for Python hosted at http://aplpy.github.com.

The STARLIGHT project is supported by the Brazilian agencies CNPq, CAPES and FAPESP and by the France-Brazil CAPES/Cofecub program.

\bibliographystyle{mnras} % style aa.bst
\bibliography{slsne.bib} % your references Yourfile.bib

%%%%%%%%%%%%%%
\appendix

\section{Stellar age estimate}
\label{appendix_sec_stellar_age}

We use the relation between EW(H$\alpha$) and stellar age of  \citet{1999ApJS..123....3L}, based of Starburst99 models. The EW-age relation depends on the assumed metallicity, $Z$, and on the initial mass function (IMF). They offer three choices of the IMF, which are similar to each other for EW(H$\alpha$) $\lesssim$ 150 $\AA$, and different metallicities, 
$Z=0.04$, 0.02, 0.008, 0.004 and 0.001.
The reference model is a power law with index $\alpha=2.35$ between low-mass and high-mass (M$_{up}$) cutoff masses of 1 and 100 $M_\odot$. 
We determined the average gas-phase oxygen abundance of the host galaxy of PTF~11hrq, $12+\log($O/H$)\sim8.15$ (see Table~\ref{tab:metallicity}). To calculate the gas-phase metallicity, Z$_{\rm gas}$, we adopt the gas-phase oxygen abundance $12+\log(O/H)$ versus gas-phase metallicity Z$_{gas}$ relationship, given in \citet{2016MNRAS.456.2140M}: 
\begin{equation}
12+\log(\mbox{O/H})=\log(Z_{\rm gas}/Z_{\odot})+9.0
\end{equation}
where Z$_{\odot}$=0.02 is the solar gas-phase metallicity.
Thus, the average gas-phase metallicity of the host galaxy of PTF~11hrq is $Z\sim0.003$. 

For comparison, if we use the solar abundance of oxygen in the log scale, $12+\log(O/H)_{\odot}$=8.69 (e.g. \citet{2009ARA&A..47..481A}) instead of 9.0, the result does not change significantly. We derive an average gas-phase metallicity of the host galaxy of PTF~11hrq $Z\sim0.006$.

\begin{figure}
\begin{center}
\includegraphics[trim=0mm 0mm 0mm 0mm, width=9cm, clip=true]{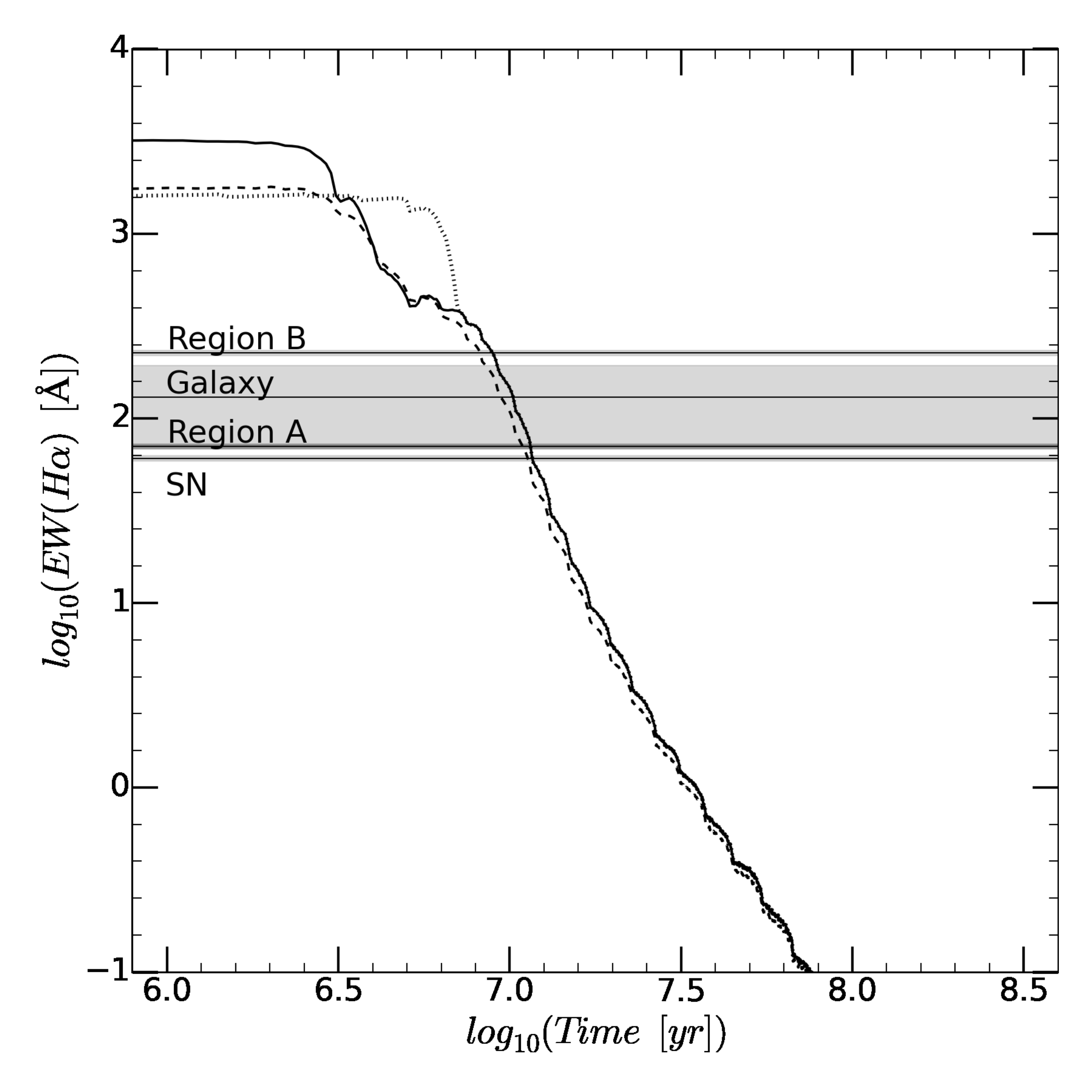}
\vspace{-4mm}
\caption{EW(H$\alpha$)-age relation for PTF~11hrq. The lines denote the Starburst99 model instantaneous star formation law with $\alpha$=2.35, M$_{up}$ = 100 M$_{\odot}$ (solid line); $\alpha$=3.30, M$_{up}$ = 100 M$_{\odot}$ (dashed line); and $\alpha$=2.35, M$_{up}$ = 30 M$_{\odot}$ (dotted line). The horizontal lines mark the average equivalent widths of the whole host galaxy, at the position of the SN and of region A and B. The gray shaded area denotes standard deviations of the average equivalent widths.}
\label{fig:age}
\end{center}
\end{figure}

Therefore, we use the Starburst99 model for the gas-phase metallicity $Z=0.004$, and estimate the stellar population age in regions A, B, and at the SN position (see Fig.~\ref{fig_MUSE_11hrq}). The results are summarized in Table~\ref{tab:metallicity}, and shown in Fig.~\ref{fig:age}.

\section{Dust extinction estimate}
\label{appendix_extinciton}

We derive the dust extinction for the host galaxy of PTF~11hrq from the Balmer decrement:
\begin{equation}
R_{\rm Balmer} =  \frac{(\mbox{H}\alpha/\mbox{H}\beta)_{\rm obs}}{(\mbox{H}\alpha/\mbox{H}\beta)_{\rm int}} ,
\end{equation}
where $(\mbox{H}\alpha/\mbox{H}\beta)_{\rm obs}$ is the observed, and $(\mbox{H}\alpha/\mbox{H}\beta)_{\rm int}$ the intrinsic $\mbox{H}\alpha/\mbox{H}\beta$ ratio. 
The $(\mbox{H}\alpha/\mbox{H}\beta)_{\rm int}$ does not strongly depend on the temperature and electron density. For example, at a fixed temperature of T=10$^4$ K, $(\mbox{H}\alpha/\mbox{H}\beta)_{\rm int}= 2.86$, 2.85 and 2.81 for $n_e = 10^2$, $10^4$, and $10^5$ cm$^{-3}$, respectively; and at a fixed electron density of $n_e = 10^4$ cm$^{-3}$, $(\mbox{H}\alpha/\mbox{H}\beta)_{\rm int}$=3.00, 2.85 and 2.74 for T=5000, 10000 and 20000 K respectively \citep[see Table 4.4. in][]{2006agna.book.....O}.

We assume $(\mbox{H}\alpha/\mbox{H}\beta)_{int}=2.86$, which is a standard choice for star forming galaxies \citep{1989agna.book.....O, 2006agna.book.....O, 2013ApJ...763..145D}. It corresponds to $T=10^4$ K and electron density $n_e = 10^2$ cm$^{-3}$.

%Because stellar absorption may be affecting the underlying continuum of the emission lines, H$\alpha$ and in particular H$\beta$, the Balmer decrement and the extinction might be overestimated. However, in our case the effect is minimal. Figure~\ref{fig_Stellarabseffect} shows the H$\beta$ line at the SN explosion site, where the stellar absorption is minimal, while at some other locations in the galaxy, e.g. the bluest region in the northern part of the galaxy, there is no visible stellar absorption.

\begin{figure}
\begin{center}
\includegraphics[trim=0mm 0mm 0mm 5mm, width=9cm, clip=true]{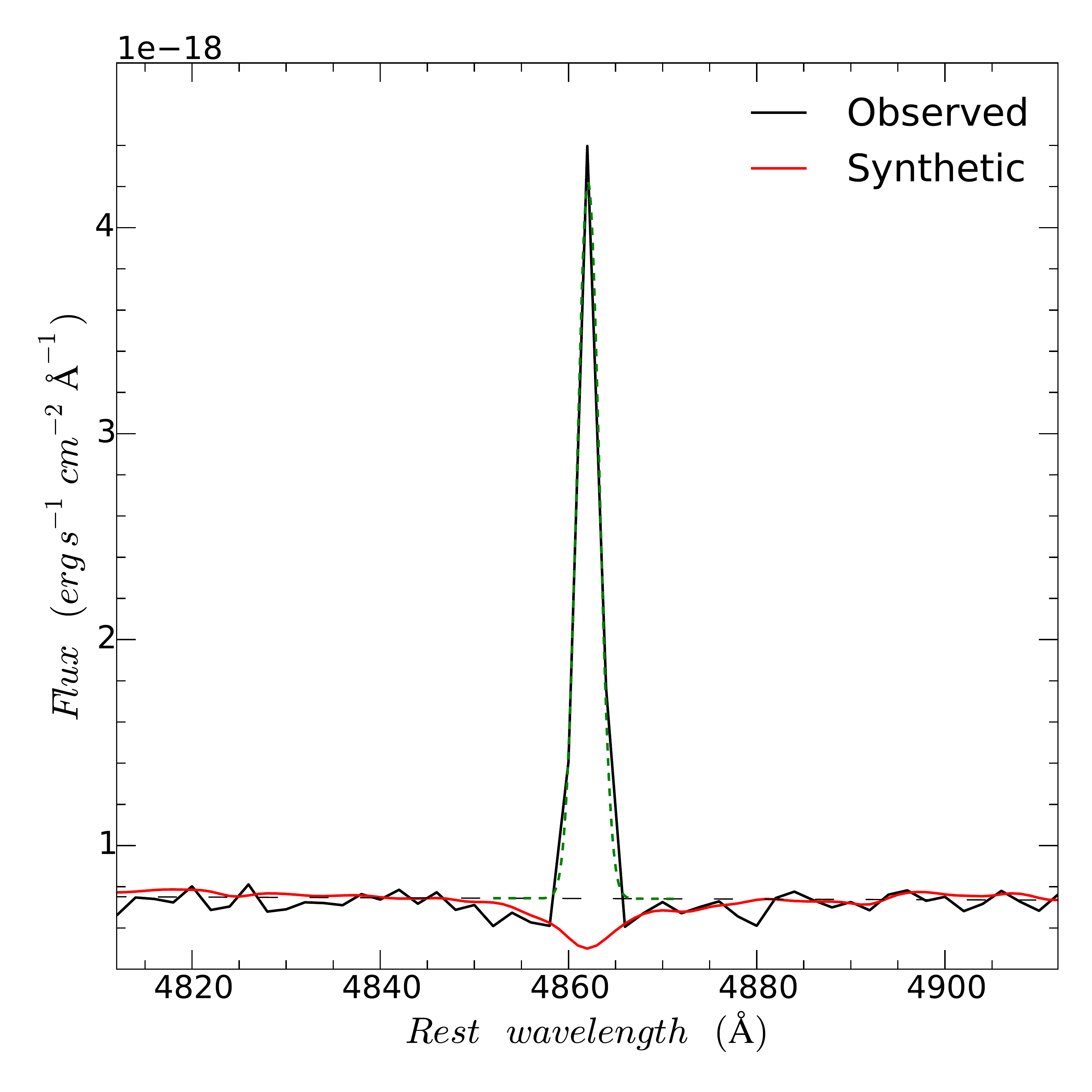}
\vspace{-4mm}
\caption{H$\beta$ line (black line) at the explosion region, and a Gaussian fit (green line) with respect to the synthetic best-fit continuum using the STARLIGHT code (red line), compared to the polynomial fit continuum (dashed line). The stellar absorption typically deducts 5-20$\%$ of the line flux.} 
\label{fig_Stellarabseffect}
\end{center}
\end{figure}

Because stellar absorption may be affecting the underlying continuum of the emission lines, H$\alpha$ and in particular H$\beta$, the Balmer decrement and the extinction might be overestimated. Figure~\ref{fig_Stellarabseffect} shows the H$\beta$ line close to the SN explosion site, where the stellar absorption is small, but not negligible.
Therefore, we use the spectral synthesis code \textsc{STARLIGHT} \citep{2005MNRAS.358..363C, 2009RMxAC..35..127C, 2011ascl.soft08006C, 2007MNRAS.381..263A, 2007MNRAS.374.1457M} to best-fit subtract the synthetic continuum before we measure the H$\alpha$ and H$\beta$ fluxes. 
\textsc{STARLIGHT} is a code that fits observed spectra (after masking the regions of known nebular emission lines, telluric absorptions, and strong night-sky emission lines) with a linear combination of a pre-defined set of  single stellar populations (SSP) base spectra of different ages and metallicities. As the base, we use 150 \citet[][, BC03]{2003MNRAS.344.1000B} SSP spectra of 6 metallicities (Z=0.0001, 0.0004, 0.004, 0.008, 0.02, 0.05) and 25 different ages ($10^6 > age_{SSP} > 18 \times 10^9$ years), which was also used in \citet{2006MNRAS.370..721M,2007MNRAS.375L..16C, 2007MNRAS.381..263A}.

\begin{figure}
\begin{center}
\includegraphics[trim=0mm 0mm 0mm 25mm, width=9cm, clip=true]{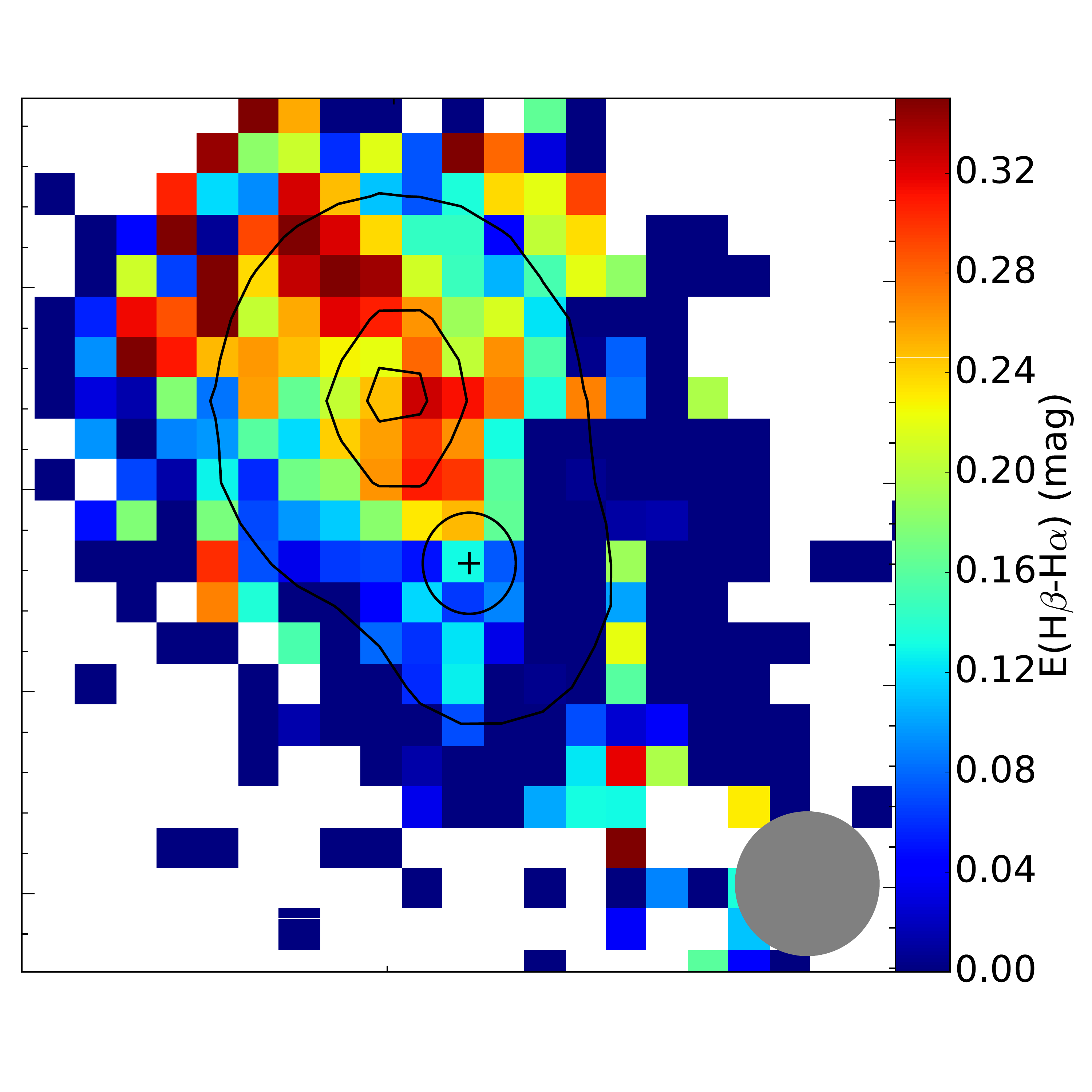}
\vspace{-12mm}
\caption{Color excess E(H$\beta$-H$\alpha$) map for the host galaxy of PTF~11hrq. The contours are same as in Fig.~\ref{fig_MUSE_11hrq}. The black ellipse denotes the SN position uncertainty.} 
\label{fig_Eba}
\end{center}
\end{figure}

The color excess can be expressed as following:
\begin{equation}
E(\mbox{H}\beta-\mbox{H}\alpha) = A(\mbox{H}\beta)-A(\mbox{H}\alpha)= -2.5 \times \log \left[  \frac{(\mbox{H}\alpha/\mbox{H}\beta)_{\rm int}}{(\mbox{H}\alpha/\mbox{H}\beta)_{\rm obs}} \right]. 
%E(\mbox{H}\beta-\mbox{H}\alpha) = A(\mbox{H}\beta)-A(\mbox{H}\alpha)= -2.5 \times \log_{10} \left[  \frac{(\mbox{H}\alpha/\mbox{H}\beta)_{\rm int}}{(\mbox{H}\alpha/\mbox{H}\beta)_{\rm obs}} \right]. 
\end{equation}
The $E(\mbox{H}\beta-\mbox{H}\alpha)$ map is shown in Fig.~\ref{fig_Eba}.

This color excess can be related to color excess in other passband filters, e.g. $E(B-V)$, or $E(\textit{F225W}-\textit{F625W})$ via an extinction law.

We adopt the \citet*{1989ApJ...345..245C} extinction law (hereafter, CCM) and calculate the ratios $E(B-V)$/$E(\mbox{H}\alpha/\mbox{H}\beta)$ and $E(\mbox{H}\beta-\mbox{H}\alpha)/E(\mbox{H}\alpha/\mbox{H}\beta)$, e.g.:
\begin{equation}
\frac{E(B-V)}{E({\rm H}\beta-{\rm H}\alpha)}=\frac{a(\frac{1}{\lambda_B})+b(\frac{1}{\lambda_B})/R_V - a(\frac{1}{\lambda_V})+b(\frac{1}{\lambda_V})/R_V}{a(\frac{1}{\lambda_{{\rm H}\beta}})+b(\frac{1}{\lambda_{{\rm H}\beta}})/R_V - a(\frac{1}{\lambda_{{\rm H}\alpha}})+b(\frac{1}{\lambda_{{\rm H}\alpha}})/R_V},
\end{equation}
where a and b are the wavelength dependent polynomials given in CCM, and $R_V$ is the total-to-selective extinction ratio, which we assume to be $R_V=3.1$. We obtain $\frac{E(B-V)}{E({\rm H}\beta-{\rm H}\alpha)}=0.93$ and $\frac{E(F225W-F625W)}{E({\rm H}\beta-{\rm H}\alpha)}=6.39$.

The CCM law is comparable to other extinction laws in the linear regime, at wavelengths $\lambda \gtrsim 0.25$ $\mu$m \citep[see][]{2003ApJ...594..279G}. 
As a sanity check, we also calculate the color excess ratios using an average extinction curve for the Small Magellanic Cloud (SMC) Wing sample \citep{2003ApJ...594..279G}, with $R_V=2.74$, and obtain $\frac{E(B-V)}{E({\rm H}\beta-{\rm H}\alpha)}=0.83$ and $\frac{E(F225W-F625W)}{E({\rm H}\beta-{\rm H}\alpha)}=5.80$. The color excess ratios calculated using the CCM extinction law are $\sim$ 1.1 times larger than with the SMC law, indicating a possible overestimation of the dust reddening with this method. 

The Galactic reddening at the position of PTF~11hrq is $E(B-V)_{\rm Gal}=0.0119 \pm 0.0005$ mag \citep{2011ApJ...737..103S}, which corresponds to  $E(\textit{F225W}-\textit{F625W})_{Gal} \simeq 0.082$ mag (assuming CCM, and $R_V=3.1$).

To calculate the reddening due to dust in the host galaxy, we subtract the Galactic reddening from the observed (total) reddening: 
%\begin{equation}
\begin{dmath}
E(\textit{F225W}-\textit{F625W})_{\rm Host}= E(\textit{F225W}-\textit{F625W})_{\rm obs}-E(\textit{F225W}-\textit{F625W})_{\rm Gal}.
\end{dmath}
%\end{equation} 
The typical value of $E(\textit{F225W}-\textit{F625W})_{Obs}$ is $0.83 \pm 0.77$ mag, and $E(\textit{F225W}-\textit{F625W})_{\rm Host}$ $\sim$ 0.8 $\pm$ 0.8 mag.

Finally, we calculate the corrected color map after resampling the E(\textit{F225W}-\textit{F625W})$_{host}$ map using \textsc{SWarp}:
%\begin{equation}
\begin{dmath}
(\textit{F225W}-\textit{F625W})_{\rm corrected}= (\textit{F225W}-\textit{F625W})_{\rm obs}-E(\textit{F225W}-\textit{F625W})_{\rm host}.
\end{dmath}
%\end{equation} 

The corrected color map is shown in Fig.~\ref{fig_ratio_11hrq}.

\section{Impact of emission lines on broad-band photometry}
\label{appendix_emissionlinesimpact}

Studies of extreme emission-line galaxies have shown that emission lines can significantly increase the brightness in broadband filters \citep[e.g.][]{Amorin2015a}. This could also affect the reported UV-to-optical colors of PTF~11hrq and PTF~12dam. In the following, we quantify the contribution of emission lines to the \textit{HST} broad-band photometry of PTF~12dam which is also the SLSN host galaxy with the strongest emission lines in the sample known today.

\begin{figure}
\begin{center}
\includegraphics[trim=0mm 0mm 0mm 0mm, width=8cm, clip=true]{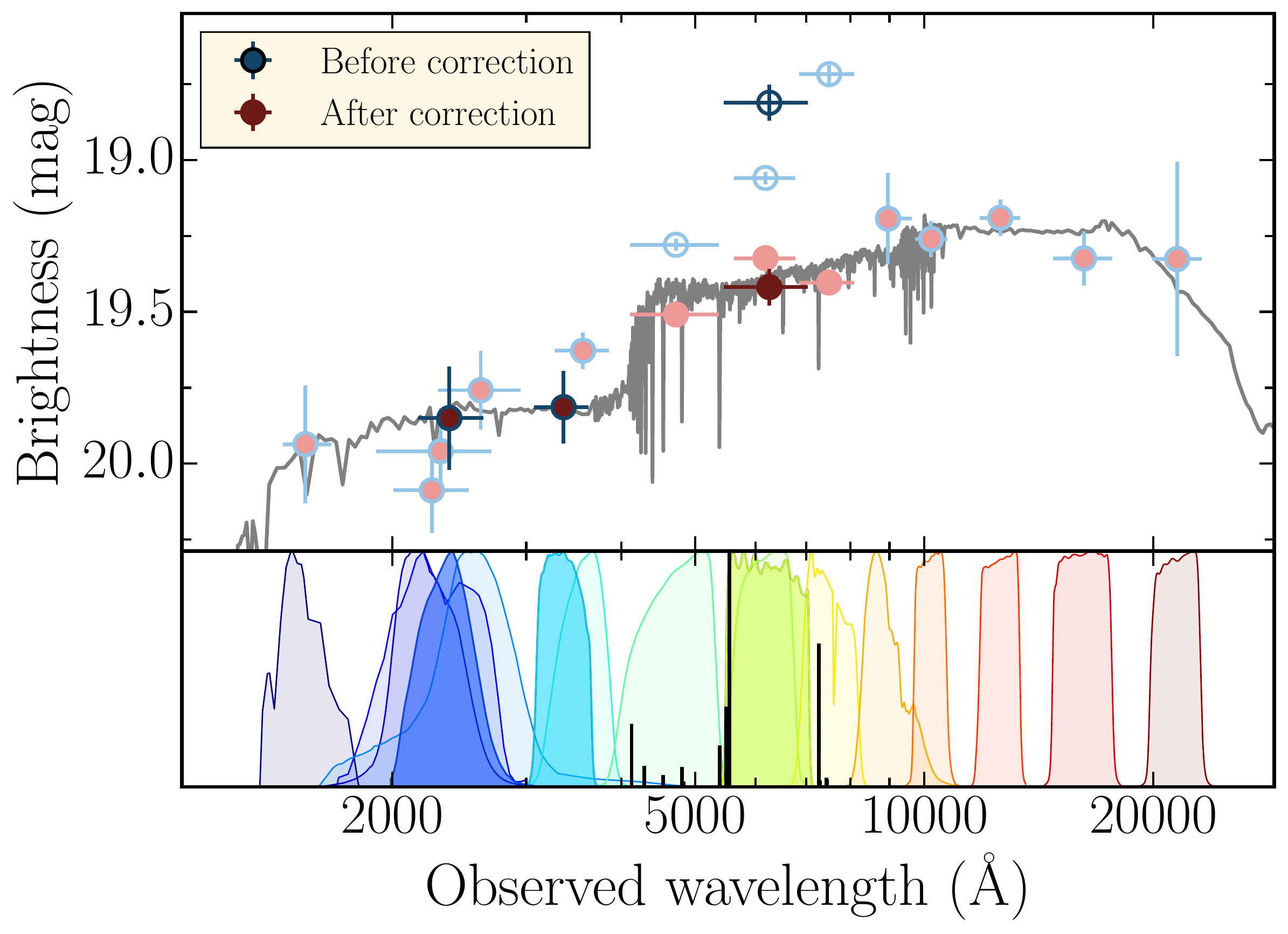}
\vspace{-1mm}
\caption{The broadband spectral energy-distribution of PTF~12dam. \textit{Top:} The empty symbols displays the photometry before the emission-line subtracting, whereas the filled symbols display the photometry after subtracting nebular emission lines. Our \textit{HST} measurements are shown in a darker shade. The gray line displays the best-fit of the SED after subtracting emission lines with a single-age stellar population (for details see \citealt{2016arXiv161205978S}). \textit{Bottom:} Synthetic emission-line spectrum built from the equivalent measurements reported in \citep{2015MNRAS.449..917L}. The shaded regions represent the transmission functions of the broad-band filters. The \textit{HST} filters are displayed in a darker tone.} 
\label{fig_photometry_impact}
\end{center}
\end{figure}

The blue empty symbols in the top panel of Fig.~\ref{fig_photometry_impact} display the photometry  presented in \citet{2016arXiv161205978S} after correcting for Galactic reddening. The new \textit{HST} photometry is shown in a darker tone. While the broad-band spectral energy distribution shows a relatively smooth evolution, the data between 4000 and 10000 \AA\, in particular the F625W photometry, show a significant excess with respect to the emission from the stellar emission (gray line in the fit).
The bottom panel shows a synthetic emission-line spectrum built from the equivalent measurements in \citet{2015MNRAS.449..917L} and the filter bandpasses corresponding to each measurement. Although [\ions{O}{III}]$\lambda\lambda$4959,5007 fall on the edge of the F625W bandpass, they still contribute $>63\%$ to the total flux.

After removing the line contribution (red data points), the broad-band spectral energy distribution is adequately fitted with a single-age stellar population (gray curve; fit taken from \citealt{2016arXiv161205978S}). This stark change in the photometry also translates into a significant bluer UV-to-optical color of $-1.98$~mag. The emission-lines of PTF~11hrq's host have significantly smaller equivalent widths resulting in a significantly smaller correction.

We note that uncertainties in the wings of the F625W filter response function could alter this correction. Furthermore, the line measurement were obtained with a long-slit observation. Spatial variations of the emission-line fluxes would manifest in a non-uniform color correction.

%%%%%%%%%%%%%

% Don't change these lines
\bsp	% typesetting comment
\label{lastpage}
\end{document}